\documentclass[aps,pre,twocolumn,floatfix,10pt]{revtex4-1}
\usepackage{graphicx}
\usepackage{dcolumn}
\usepackage{bm}
\usepackage{latexsym}
\usepackage{array,booktabs}
\newcolumntype{L}{@{}>{\kern\tabcolsep}l<{\kern\tabcolsep}}
\usepackage{colortbl}
\usepackage{xcolor}
\usepackage{amsmath}

\newcommand{\RN}[1]{%
\textup{\uppercase\expandafter{\romannumeral#1}}%
}

\begin{document}
\title{{\bf A Multispecies Exclusion Process with Fusion and Fission of Rods:\\ a model inspired by Intraflagellar Transport }}
\author{Swayamshree Patra} 
\affiliation{Department of Physics, Indian
  Institute of Technology Kanpur, 208016, India} 
\author{Debashish Chowdhury{\footnote{Corresponding author; e-mail: debch@iitk.ac.in}}}
\affiliation{Department of Physics, Indian Institute of Technology
  Kanpur, 208016, India}
\begin{abstract}
\noindent {\bf Abstract}: 
We introduce a multispecies exclusion model where length-conserving probabilistic fusion and fission of the hard rods are allowed. Although all rods enter the system with the same initial length ${\ell}=1$, their length can keep changing, because of fusion and fission, as they move in a step-by-step manner towards the exit. Two neighboring hard rods of lengths ${\ell}_{1}$ and ${\ell}_{2}$ can fuse into a single rod of longer length ${\ell}={\ell}_{1}+{\ell}_{2}$ provided ${\ell} \leq N$. Similarly, length-conserving fission of a rod of length ${\ell}' \leq N$ results in two shorter daughter rods. Based on the extremum current hypothesis, we plot the phase diagram of the model under open boundary conditions utilizing the results derived for the same model under periodic boundary condition using mean-field approximation. The density profile and the flux profile of rods are in excellent agreement with computer simulations. Although the fusion and fission of the rods are motivated by similar phenomena observed in Intraflagellar Transport (IFT) in eukaryotic flagella, this exclusion model is too simple to account for the quantitative experimental data for any specific organism. Nevertheless, the concepts of `flux profile' and `transition zone' that emerge from the interplay of fusion and fission in this model are likely to have important implications for IFT and for other similar transport phenomena  in long cell protrusions.
\end{abstract}

\maketitle

\section{introduction}

Non-equilibrium stationary state (NESS) of a driven system \cite{blythe07,evans05,bennaim10} is the counterpart of the state of equilibrium in a thermodynamic system. Totally asymmetric simple exclusion process (TASEP) \cite{schutz01,derrida98,mallick15} is a paradigmatic model for theoretical studies of the fundamental physical principles underlying NESS in systems of interacting self-driven particles. 

A non-vanishing average uni-directional flow of particles through the system in its stationary state is a {\it macroscopic} indicator of the fact that steady state of TASEP is never in equilibrium. Under open boundary conditions (OBC), the density profile, which depicts the average stationary site occupational probabilities, is another {\it macroscopic} characteristic of the NESS of TASEP. Both these fundamental macroscopic characteristics of the NESS of TASEP can be computed, as averages, from the stationary configurational probabilities $\{P^{ss}(C)\}$ and the corresponding probability currents $\{J^{ss}(C \to C')\}$, which together provide a {\it complete and unique microscopic} description of each NESS of TASEP \cite{zia07}.

In more general formulations of TASEP, the particles are replaced by hard rods, each of length ${\ell}$, where the length of the rods are measured in the units of lattice spacing \cite{schonherr04,dong07,mishra17}. From now onwards we will refer to particles also as rods with ${\ell}=1$. Multi-species TASEP are known to exhibit richer varieties of phenomena compared to those in single-species TASEP \cite{schutz03}. Both single-species and multi-species TASEP, and their various extensions, have also found applications in modeling collective phenomena at many scales, starting from macroscopic vehicular traffic on highways to molecular motor traffic on filamentous tracks in living cells \cite{lipowsky01,parmeggiani03,css00,chou11,rolland15,lipowsky06,muhuri10,greulich12,parmeggiani14,sugden07,evans11,dong12,pinkoviezky14,graf17}.

The distinct species of rods can be distinguished by either their length or their distinct kinetics (or both).  In this paper we study a biologically motivated exclusion process with $N$ ($N > 1$) allowed species of rods, ${\ell}$-th species having length ${\ell}$ (in the units of lattice spacing), where the species are {\it interconvertible} because of the ongoing {\it fusion} and {\it fission} of the rods. Two rods of length ${\ell}'$ and ${\ell}''$ (${\ell}',{\ell}''$=1,2,3..N) in contact with each other are allowed to fuse resulting in a longer rod of length ${\ell}={\ell}'+{\ell}''$, provided ${\ell} \leq N$. Similarly, a rod of length ${\ell}$ (${\ell} \leq N$) can split into two shorter rods of lengths ${\ell}'''$ and ${\ell''''}$. 
The constraint imposed on the maximum size of a rod can be relaxed by allowing the limit $ N \to \infty$.

The model developed in this paper is motivated by 
intraflagellar transport (IFT), which is directed stochastic transport of molecular cargoes in long protrusions of some eukaryotic cells. A brief summary of IFT is presented in the next section. 
Although the processes of fusion and fission of the rods in our model is motivated by IFT, this model is not intended to account for experimental data in any specific flagellated cell. Instead, the model focusses on the consequences of ongoing fusion and fission on the collective spatio-temporal organization of the $N$ species of interconvertible particles in the NESS of the system.

By a combination of mean-field theory (MFT) and Monte Carlo (MC) simulations, we demonstrate qualitatively distinct features of the density profile and flux in this model. The density profile exhibits a ``transition zone'' (TZ) whose thickness depends on the fusion-fission kinetics. We introduce the concept of ``flux profile'' to highlight the relative contributions of the interconvertible species of particles and rods to the overall flux as they move forward along the lattice.

\section{Brief introduction to IFT}

Transport of various types of molecular and membrane-bound cargoes in eukaryotic cells is carried out by molecular motors that are driven along filamentous tracks \cite{chowdhury13,kolomeisky15}. Tubular stiff filaments, called microtubule (MT) serve as tracks for two `superfamilies' of molecular motors which move naturally in opposite directions by consuming chemical fuels. Cargo transport plays a crucial role in the growth, maintenance and shrinkage of wide varieties of long protrusions of cells. A {\it flagellum} is a membrane-bound cylindrical cell protrusion found in some eukaryotic cells (for example, unicellular  eukaryote  {\it Chlamydomonas reinhardtii}). 
Inside this cylinder nine doublet MTs, arranged in a cylindrically symmetric fashion, extend from the base to the tip of the protrusion. The eukaryotic flagella (not to be confused with bacterial flagella) are also referred to as cilia. 

IFT is the phenomenon of bidirectional transport of multi-subunit protein complexes, called IFT particles, within the space between the MT and the ciliary membrane, where the motors hauling the cargoes walk along the MT tracks. 
Because of their superficial similarities with cargo trains hauled along railway tracks, chain-like assemblies formed by IFT particles are called IFT trains  \cite{ishikawa17,buisson12,stepanek16,lechtrek17}.

 In principle, stochastic stepping of motors can speed up a following IFT train or slow down a leading IFT train thereby causing their physical contact that, occasionally, leads to their probable fusion. Similarly, abrupt tension generation by asynchronous stepping of motors can, in principle, rupture the bond between two neighboring IFT particles in an IFT train which manifests as a probabilistic fission event. Fusion and fission of IFT trains have, indeed,  been observed \cite{stepanek16,buisson12,lechtrek17}. These stochastic fusion and fission processes are captured by the model reported in this paper.

Any model intended to account for experimental data on IFT must describe both anterograde (tipward) and retrograde (baseward) IFT within a single theoretical framework. Therefore, it has to include not only two distinct tracks, representing a MT doublet, but also allow different values of the model parameters for the kinetics of anterograde and retrograde IFT trains. Inadequacy of experimentally measured quantitative data from the same flagellated cell makes it difficult to assign these rates. Moreover, the effects of the observed tight association between the inner surface of the ciliary membranes and the IFT trains on the structure and dynamics of the latter is not known and, hence, difficult to model theoretically. Furthermore, complete specification of such a model of IFT will also require  prescriptions for coupling the anterograde and retrograde fluxes at the flagellar tip as well as at the base; experimental information available at present are not adequate to prescribe such rules \cite{wingfield17}. Therefore, no attempt is made in this paper to develope a complete kinetic model that would account for experimental data on IFT in any specific flagellated cell.

\begin{figure}[ht]
\begin{center}
\includegraphics[width=1.0\columnwidth]{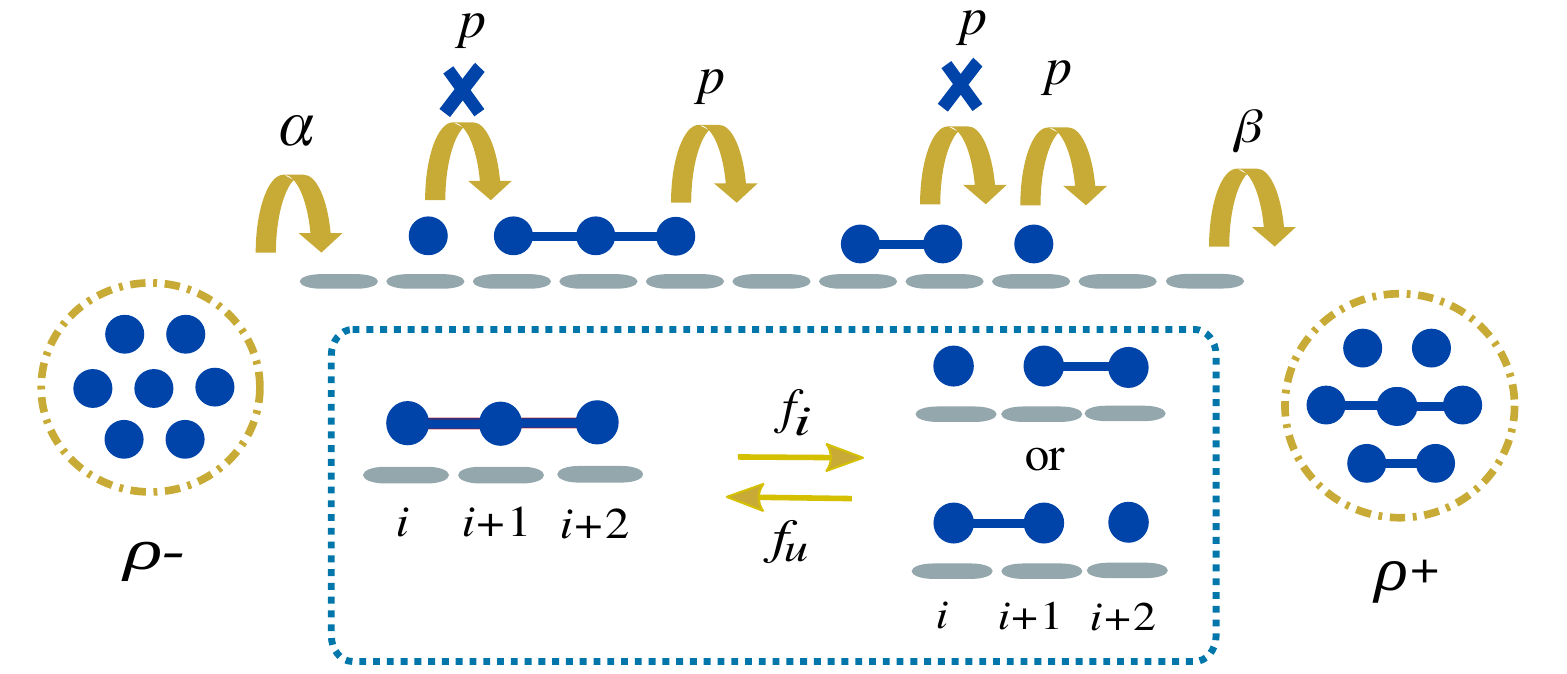}
\end{center}
\caption{(Color online) {\bf{Model}}: Only rods of length $\ell=1$ enter the 1-D lattice from the left. Such a rod can occupy the leftmost site, with rate $\alpha$, only if it is not already occupied by another at that instant of time. Then, obeying exclusion, the rods hop forward with length-independent hopping rate $p$.  In addition, two neighbouring rods can undergo fusion, with rate $f_u$, thereby resulting in a single rod of length ${\ell}$, provided ${\ell} \leq N$. Any rod of length ${\ell} > 1$ can suffer a fission, with rate $f_i$, resulting in two neighbouring rods. The rods exit, with rate 
$\beta$, from the last site.Note that each fusion and fission event conserves total length (equivalently, mass).  All possible combinations of the pairs that conserve the total length are equally probable result of a fission event. The probability of fission and that of exit from the last site are both independent of the instantaneous length of a rod. }
\label{fig1}
\end{figure}

\section{Model and methods} 

The track is denoted by a one-dimensional lattice where each site is labelled by the integer index $i(i=1,..,L)$; the sites $i=1$ and $i=L$ correspond to the sites of entrance and the exit, respectively, at the two boundaries of the lattice  (see Fig.\ref{fig1}). {\it Location} of the rod is specified by the lattice site $i$ which the {\it leftmost} tip of the rod occupies. The probability of finding a rod of length $\ell$ at site $i$ at time $t$ is denoted by $P_{\ell}(i,t)$; in a steady state of the system, $P_{\ell}(i,t)$ becomes independent of time $t$. A rod of length $\ell$ at site $i$ covers site $i$ to $i+\ell-1$ simultaneously, with its leftmost tip at site $i$. Because of  mutual exclusion, none of the sites can be covered simultaneously by more than one rod. The mutual exclusion is captured by the conditional probability  $\xi(\underline{i}|i+\ell)$ that the site $i+\ell$ is not covered by another rod, given that there is a rod of length $\ell$ at site $i$ (see appendix for derivation).

Only rods of length $\ell=1$ enter the 1-D lattice from left and occupy the leftmost site ($i=1$), with rate $\alpha$, provided that site is not already occupied at that instant by any other rod.  After entry, obeying exclusion, the rods hop forward with length-independent hopping rate $p$. 

Two neighbouring rods of length $\ell'$ and $\ell''$ can fuse, with rate $f_u$, resulting in a single rod of length 
$\ell={\ell}'+{\ell}''$ provided ${\ell} \leq N$. Similarly, any rod of length ${\ell} > 1$ can split, with rate $f_i$, resulting in two neighbouring particles of lengths ${\ell}'''$ and ${\ell}''''$ such that ${\ell}'''+{\ell}''''={\ell}$. Thus, each fusion and fission event is a length-conserving process. All the rods of length ${\ell} > 1$ are equally prone to fission irrespective of their individual instantaneous lengths; all possible pairs $\{{\ell}', {\ell}''\}$ that satisfy ${\ell}' + {\ell}'' = {\ell}$ are equally probable result of a fission event. For convenience, we define a dimensionless 'stickiness' parameter $K=f_u/f_i$. Some other possible alternative rules for the fusion-fission kinetics of the rods, mentioned in the concluding section, will be explored in a future publication \cite{patra18}.

The rods of all lengths ${\ell}$ (${\ell}=1,2,...,N$) exit from right boundary ($i=L$) with a length-independent rate $\beta$. There are $\ell-1$ dummy sites available beyond $i=L$ (from $i=L+1$ to $i=L+\ell-1$). For rods located at $i>(L-\ell)$), the leading edge is out of the track (resting on the dummy sites), and these rods can hop from $i$ to $i+1$ with rate $p$ without any hindrance as dummy sites are always available. 

The entry and exit of rods described above are essential for a complete specification of the kinetics of the model under open boundary conditions (OBC). However, by converting the finite lattice into a closed chain one can reduce the model to a simpler version with periodic boundary condition (PBC). Analysing the system under PBC gives insights into the interplay of forward hopping and fusion-fission processes. Moreover, as we show later, important results for the model under the more realistic OBC can be derived exploiting the results obtained under PBC. 

In the NESS the fraction of each species (which is identical to their respective probabilities), for a given $K$, is given by {\small $ \mathcal{F}_{\ell}={P_{\ell}}/{\sum_{\ell=1}^N P_{\ell}}$} from which the average length $\langle {\ell} \rangle$ and the randomness parameter $\mathcal{R}$, which is a measure of the length fluctuations, can be calculated using
\begin{equation}
    \langle {\ell} \rangle=\sum_{\ell=1}^{N} {\ell}~{\mathcal{F}_{\ell}} \ \ \ \ \ \ \ \ \mathcal{R}=\frac{\sqrt{\sum_{\ell=1}^N(\ell-\langle\ell\rangle)^2 ~\mathcal{F}_{\ell}}}{\langle\ell\rangle}
\end{equation}

The ${\ell}$-dependent number flux $\mathcal{J}_{\ell}$  is defined as the number of rods of length $\ell$ passing through a given point per unit time.  The mass flux $\mathcal{J}_{mass}$ is defined as the total mass of the rods (in our units, a rod of length $\ell$ is assumed to have a mass ${\ell}$) which passes through a given point per unit time.  Thus, these two fluxes are given by
\begin{equation}
    \mathcal{J}_{\ell}=p~{P_{\ell}}~\xi(\underline{i}|i+\ell) ; \  \   \ ~{\rm and}~  \  \  \  \   \  \mathcal{J}_{mass}=\sum_{\ell=1}^N \ell ~  \mathcal{J}_{\ell}
\end{equation}
respectively.


For a theoretical treatment of the model, the exact multi-site configurational probabilities are approximated, under mean-field approximation (MFA), by products of single-site occupational probabilities $P_{\ell}(i,t)$ ($1 \leq i \leq L$). 
Master equations governing the time evolution of the probabilities $P_{\ell}(i,t)$ ($1 \leq i \leq L$) are written down capturing all the kinetic processes, namely, entry, exit, hopping, fusion and fission of the rods. 

In the NESS, under PBC, these master equations are independent not only of time $t$ but also of the site index $i$. Because of this additional simplicity, the equations can be solved analytically to derive the corresponding expressions for the characteristic quantities introduced above. However, the master equations remain site-dependent even in the NESS under OBC; consequently, treatments under OBC require combination of analytical and numerical solutions as described in the sections below where these results are presented.

In the MC simulation of our model, the lattice sites, which are denoted by the integers $i$ ($1 \leq i \leq L$), are chosen randomly with equal probability. The status of the site is then updated according to the kinetics that defines the model. In other words, we implement a site-oriented random-sequential updating rule. A total of $L$ successive updates constitutes a single MC step. Since we are interested exclusively in the steady state of the system, it is adequate to express time in the units of MC steps. Accordingly, all the rates $p, \alpha, \beta, f_{u}, f_{i}$ are appropriately converted to the corresponding probabilities per MC step (MCS). 

If the randomly chosen site $i$ happens to accommodate the leftmost tip of a rod of length ${\ell}$, then one of the three possible mutually exclusive choices for updating is implemented according to the following rules: \\
(i) with the probability $p/(2f_u+f_i+p)$ the the rod hops forward, provided the downstream site $(i+\ell)$ is empty; 
(ii) with the probability $f_{u}/(2f_u+f_i+p)$ the rod fuses with the adjacent rod of length ${\ell}'$ touching its front (or, with the same probability, the rod fuses with the neighboring rod of length ${\ell}'$ touching its rear) provided 
$\ell'+ \ell \leq N$; 
(iii) with the probability $f_{i}/(2f_u+f_i+p)$ a fission of the rod occurs where all  possible pairs $\{{\ell}', {\ell}''\}$ that satisfy ${\ell}' + {\ell}'' = {\ell}$ are equally probable result of a fission event. \\ 
For the case of OBC, additional update rules need to specified for the entry and exit at the left and right boundaries, located at $i=1$ and $i=L$, respectively. At the left boundary of the lattice, entry of only a particle of length ${\ell}=1$ is allowed, with rate $\alpha$, if and only if the site $i=1$ is empty. At the right boundary, if the left edge of a rod is located at $i=L$, it is no more allowed to fuse or split, irrespective of its length, but it can make an exit from the lattice with a length-independent rate $\beta$. 

As the state system was updated following the rules listed above, the flux profile was monitored continuously. Long before the completion of the first $10^7$ MCS, the flux profile became steady (except for minor fluctuations around the steady value) thereby indicating attainment of the NESS. Thereafter the numerical data for the various properties of interest were collected for the next $10^7$ MCS to compute the average steady-state properties.

\section{Results under PBC} 

We first present the master equations for an arbitrary value of the integer N. As described above, the probability of finding a rod of length $\ell$ at site $i$ at time $t$ is denoted by $P_{\ell}(i,t)$. Time evolution of $P_{\ell}(i,t)$ ($1 \leq {\ell} \leq N$) are given by the following master equations

\begin{widetext}
\begin{eqnarray}
&\frac{dP_{\ell}(i,t)}{dt} =\overbrace{pP_\ell(i-1,t)\xi_{N}(\underline{i-1}|i+{\ell}-1)-pP_\ell(i,t)\xi_{N}(\underline{i}|i+{\ell})}^{\text{HOPPING terms}}
&\nonumber\\  \nonumber\\
&+\overbrace{\underbrace{f_i{\sum_{s={\ell}+1}^{N}}{(\frac{1}{s-1})P_{s}(i,t)}}_{\substack{ \text{fission of a rod of length $s(>\ell)$ } \\ \text{ located at $i$} }}+\underbrace{f_i{\sum_{s={\ell}+1}^{N}}{(\frac{1}{s-1})P_s(i-s+{\ell},t)}}_{\substack{ \text{fission of a rod of length $s(>\ell)$ } \\ \text{ located at $i-s$}}}}^{\text {gain by FISSION}}+\overbrace{\underbrace{{f_u}{\sum_{s=1}^{{\ell}-1}}P_{s}(i)P_{{\ell}-s}(i+s)}_{\substack{\text{fusion of a rod of length $s$ at $i$ } \\ \text {with a rod of length $\ell-s$ at $i+s$}}}}^{\text{gain by FUSION}}& \nonumber\\ \nonumber\\
&-\overbrace{\underbrace{f_i{P_{\ell}(i,t)}}_{\text{fission of a rod of length $\ell$ located at $i$}}}^{\text {loss by FISSION}}-\overbrace{\underbrace{{f_u}{\sum_{s=1}^{N-\ell}P_{\ell}(i,t)P_s(i+{\ell},t)}}_{\substack{\text{fusion of rod of length $\ell$ at $i$ } \\ \text{ with rod of length $s$ at $i+\ell$}}}- \underbrace{{f_u}{\sum_{s=1}^{N-\ell}P_{\ell}(i,t)P_{s}(i-s,t)}}_{\substack {\text{fusion of rod of length $\ell$  } \\ \text{ at $i$ with rod of length $s$ at $i-s$}}}}^{\text {loss by FUSION}}
\label{master_eq_gen}
\end{eqnarray}

for all $i$ ($1 \leq i \leq L$). 
Although the equations are written above for an arbitrary $N > 1$, for simplicity, we present our analysis only for the special cases N=2 and  N=3.
 Master-equations governing the evolution of rods for N=2 are 
\begin{eqnarray}
\frac{dP_1(i,t)}{dt}=\underbrace{ pP_1(i-1){\xi}_2-pP_1(i){\xi}_2}_{\text {hopping terms}} +\underbrace{f_iP_2(i,t)+f_iP_2(i-1,t)}_{\text {gain by fission of $\ell=2$ }}\nonumber\\\underbrace{-f_uP_1(i)P_1(i+1)}_{\text {loss by fusing with a rod stalled ahead \ \ \ \ \ \ \ }}  \underbrace{-f_uP_1(i)P_1(i-1)}_{\text {loss by fusing with a rod stalled behind}} 
\label{master_eq_P1}
\end{eqnarray}
\begin{eqnarray}
\frac{dP_2(i,t)}{dt}=\underbrace{pP_2(i-1){\xi_2}-pP_2(i){\xi_2}}_{\text{hopping terms}} \ \
+ \ \underbrace{f_uP_1(i)P_1(i+1)}_{\text {gain by fusion of $\ell=1$}}-\underbrace{f_iP_2(i,t)}_{\text{loss by fission}}
\label{master_eq_P2}
\end{eqnarray}

Master-equation governing the evolution of rods for N=3 are 
\begin{eqnarray}
\frac{dP_1(i)}{dt}=&\underbrace{pP_1(i-1)\xi_3 -pP_1(i)\xi_3 }_{\text {hopping terms}} +\underbrace{f_iP_2(i)+f_iP_2(i-1)+\frac{f_i}{2}P_3(i)+\frac{f_i}{2}P_3(i-2)}_\text{gain by fission of $\ell=2$ and $\ell=3$}\nonumber\\& 
\underbrace{{-f_uP_1(i)P_1(i+1)-f_uP_1(i)P_1(i-1)}}_\text{loss by fusing with a rod }
\underbrace{{-f_uP_1(i)P_2(i+1)-f_uP_1(i)P_2(i-1)}}_\text{loss by fusing with a rod }
\end{eqnarray}
\begin{eqnarray}
\frac{dP_2(i)}{dt}=&\underbrace{pP_2(i-1)\xi_3-pP_2(i)\xi_3}_{\text {hopping terms}}+\underbrace{{\frac{f_i}{2}P_3(i)}+{\frac{f_i}{2}P_3(i-1)}}_{\text {gain by fission of $\ell=3$}}+\underbrace{f_uP_1(i)P_(i+1)}_{\text {gain by fusion }}\nonumber\\&-\underbrace{f_uP_2(i)P_1(i+2)-f_uP_2(i)P_1(i-1)}_{\text {loss by fusion}}-\underbrace{f_iP_2(i)}_{\text{loss by fission}}
\end{eqnarray}
and
\begin{eqnarray}
\frac{dP_3(i)}{dt}=&\underbrace{pP_3(i-1)\xi_3 -pP_3(i)\xi_3}_{\text {hopping terms}}\underbrace{-f_iP_3(i)}_{\text {loss by fission}}
+\underbrace{f_uP_1(i)P_2(i+1)+f_uP_2(i)P_1(i+2)}_{\text {gain by fusion}}
\end{eqnarray}
\end{widetext}

\begin{figure}[ht]
\begin{center}
\includegraphics[width=1.0\columnwidth]{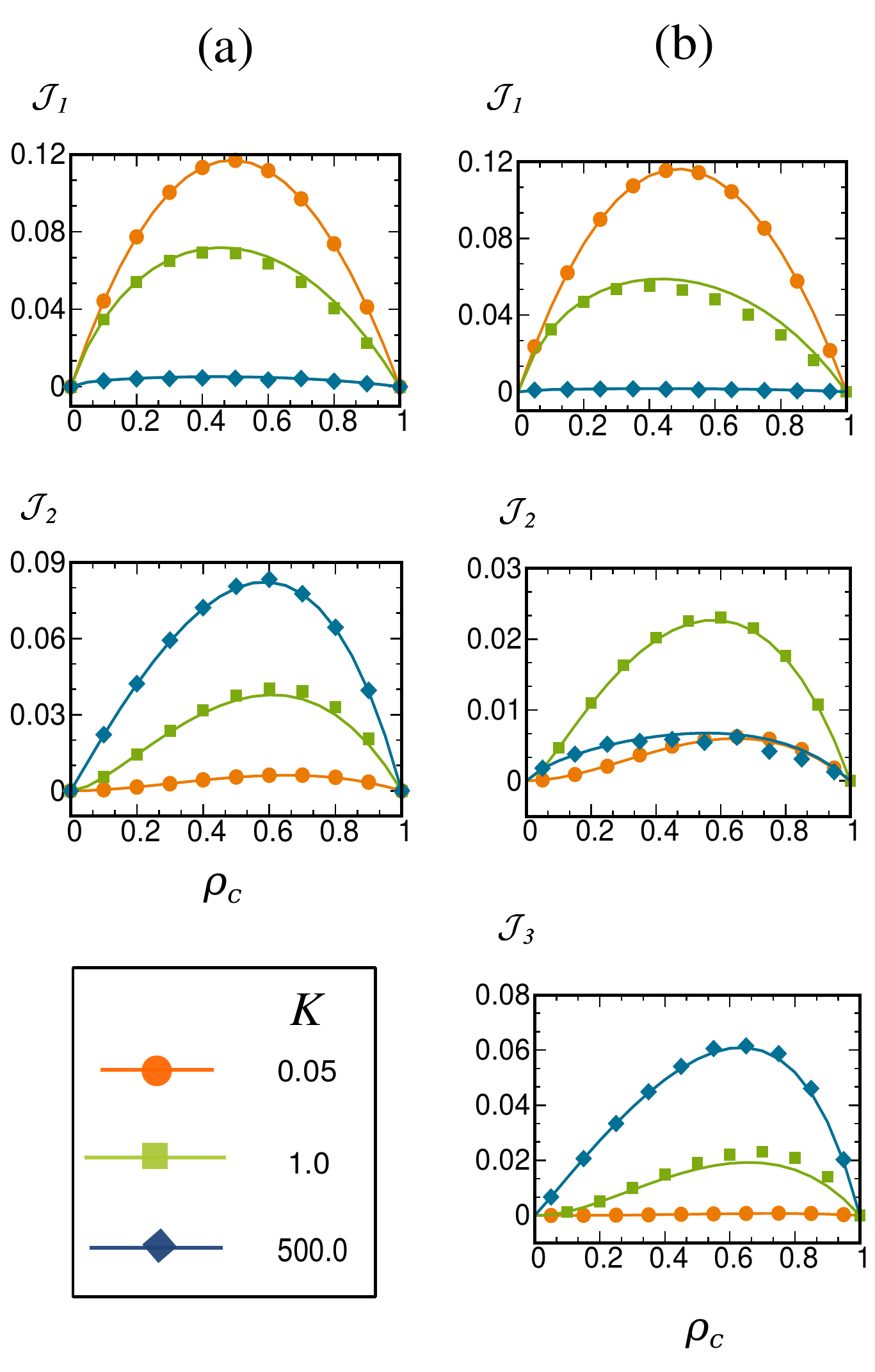}
\end{center}
\caption{ (Color online) Species specific Flux Profiles under PBC for different values of stickiness $K$ for (a) N=2 and (b) N=3 (Lines--MFT ; Dots--Simulation) }
\label{fig2}
\end{figure}

\begin{figure}[ht]
\begin{center}
\includegraphics[width=1.0\columnwidth]{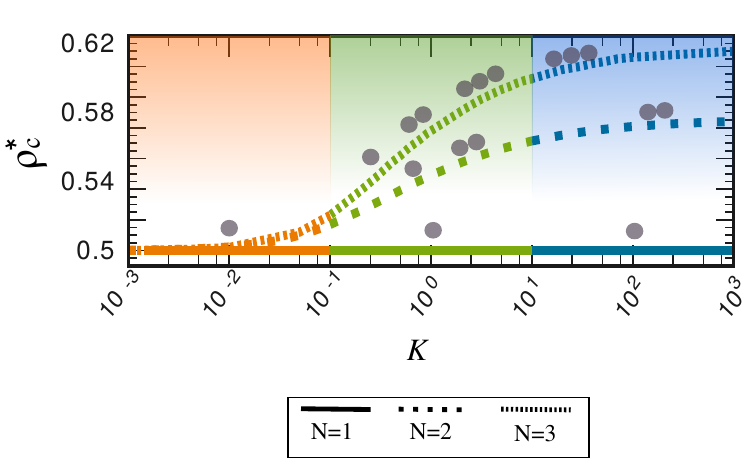}
\end{center}
\caption{(Color online) The coverage density $\rho_C^*$, that corresponds to the maximum of  $\mathcal{J}_{mass}$ obtained in mean-field theory under PBC, are plotted against the stickiness $K$ for $N=1,2,3$. The number of contiguous bullets indicate the rod sizes.}
\label{fig3}
\end{figure}

By definition, in the stationary state the probabilities become independent of time, i.e., $dP_{\ell}/dt$=0. Moreover, because of the translation symmetry of the stationary state under PBC, the site dependence of $P_{\ell}(i)$ also drops out, i.e., $P_{\ell}(i)=P_{\ell}$ for all $i$. Therefore, in the stationary state, the master equations for the system under PBC, written in MFA, reduce to   
\begin{equation}
KP_1^2-P_2=0
\end{equation}
for N=2, subjected to the constraint
\begin{equation}
P_1+2P_2=\rho_c
\end{equation} 
while those for  N=3 reduce to 
\begin{eqnarray}
2KP_1P_2-P_3=0\nonumber\\
KP_1^2-P_2=0
\end{eqnarray}
subjected to the constraint
\begin{equation}
P_1+2P_2+3P_3=\rho_c.
\end{equation} 
Thus, under PBC, system of $N$ master equations for the model reduce to $N-1$ equations, recast in terms of $K$,
subjected to the following general mass conservation constraint
\begin{equation}
\sum_{\ell=1}^{N} \ell P_{\ell}=\rho_c
\end{equation}
In Table I, all the $\ell$-dependent quantities  and in Table {\ref{table-2}} some other quantities are summarised for N=2,3. Number densities and other quantities are expressed in compact form using the terms $\zeta_{N,n}$ (n=1,2,..,N-1) which are the functions of $\rho_c$ and $K$.\\

\begin{widetext}


\begin{table}
 \label{table-1}
  \centering
  \caption{\bf \large $\ell$-dependent Quantities For N=2 and N=3}

\begin{tabular}
{@{} c L  L | L |  L |  L  @{} >{\kern\tabcolsep}c @{} @{} >{\kern\tabcolsep}c @{}  @{} >{\kern\tabcolsep}c @{} @{} >{\kern\tabcolsep}c @{}}
\hline
\hline
 & &  & &   &   \\ 
{\bf {\large N}} & &  &  {\bf{\large ${\ell}=1$}}  &  {\bf{\large ${\ell}=2$}}   & {\bf{\large${\ell}=3$}}  \\ 
 & &  & &   &   \\ 
\hline
\hline
\\

\text{\bf {\large Number Density ($\mathcal{P}_{\ell}$)  }}    \\\midrule
\hline
\hline
\rowcolor{black!10}[0pt][0pt]  &  &  &  &  &\\
\rowcolor{black!10}[0pt][0pt] {\large N=2} & &  & {\large $\zeta_{2,1}$ }& {\large $\frac{1}{2}({\rho_c-\zeta_{2,1}})$ }& \\
\rowcolor{black!10}[0pt][0pt]  &  &  &  &  &\\
\rowcolor{black!20}[0pt][0pt]  &  &  &  &  &\\
\rowcolor{black!20}[0pt][0pt] {\large  N=3} & &  & {\large $\zeta_{3,1}$} & {\large $\zeta_{3,2}-\frac{2}{9}\zeta_{3,1}$} & {\large $\frac{\rho_c}{3}-\frac{5}{27}\zeta_{3,1}-\frac{2}{3}\zeta_{3,2}$}\\
\rowcolor{black!20}[0pt][0pt]  &  &  &  &  &\\
\hline
\text{\bf                     }    \\\midrule
\text{\bf {\large  Number Flux ($\mathcal{J}_{\ell}$)  } }   \\\midrule
\hline
\hline
\rowcolor{black!10}[0pt][0pt]  &  &  &  &  &\\
\rowcolor{black!10}[0pt][0pt] {\large  N=2} & &  & {\Large  $\frac{2p(1-\rho_c )\zeta_{2,1}}{(2-\rho_c)+\zeta_{2,1}}$ } & {\Large  $\frac{p(1-\rho_c )(\rho_c-\zeta_{2,1})}{(2-\rho_c)+\zeta_{2,1}}$ }&    \\
\rowcolor{black!10}[0pt][0pt]  &  &  &  &  &\\
\rowcolor{black!20}[0pt][0pt]  &  &  &  &  &\\
\rowcolor{black!20}[0pt][0pt] {\large N=3} & &  &{\Large   $\frac{27p(1-\rho_c)\zeta_{3,1}}{16\zeta_{3,1}+9(3+2\zeta_{3,2}-2\rho_c)}$ } & {\Large $\frac{3p(1-\rho_c)(9\zeta_{3,2}-2\zeta_{3,1})}{16\zeta_{3,1}+9(3+2\zeta_{3,2}-2\rho_c)}$ } & {\Large $\frac{p(1-\rho_c)(9\rho_c-5\zeta_{3,1}-18\zeta_{3,2})}{16\zeta_{3,1}+9(3+2\zeta_{3,2}-2\rho_c)}$}
 \\
\rowcolor{black!20}[0pt][0pt]  &  &  &  &  &\\
\hline
\text{\bf     }    \\\midrule
 \text{\bf {\large Fraction ($\mathcal{F}_{\ell}$)    }   }      \\\midrule
\hline
\hline
\rowcolor{black!10}[0pt][0pt]  &  &  &  &  &\\
\rowcolor{black!10}[0pt][0pt] {\large  N=2} & &  & {\Large  $\frac{2 \zeta_{2,1}}{{\rho_c + \zeta_{2,1}}}$ } & {\Large  $\frac{\rho_c - \zeta_{2,1}}{{\rho_c + \zeta_{2,1}}}$ }&  \\ 
\rowcolor{black!10}[0pt][0pt]  &  &  &  &  &\\
\rowcolor{black!20}[0pt][0pt]  &  &  &  &  & \\
\rowcolor{black!20}[0pt][0pt] {\large  N=3} & &  & {\Large $\frac{27 \zeta_{3,1}}{9 \rho_c + 16 \zeta_{3,1} + 9 \zeta_{3,2}}$} & {\Large $\frac{27 \zeta_{3,2}-6\zeta_{3,1}}{9 \rho_c + 16 \zeta_{3,1} + 9 \zeta_{3,2}}$} & {\Large  $\frac{9 \rho_c - 5 \zeta_{3,1} - 18 \zeta_{3,2}}{9 \rho_c + 16 \zeta_{3,1} + 9 \zeta_{3,2}}$}\\ 
\rowcolor{black!20}[0pt][0pt]  &  &  &  &  & \\

\bottomrule
 \hline
\end{tabular}
\end{table}

These are


\begin{equation}
\zeta_{2,1}=\frac{\sqrt{1+8 K \rho_c}-1}{4K}
\end{equation}

\begin{eqnarray}
\zeta_{3,1}=-\frac{1}{9 K}-\frac{7}{9\ 2^{1/3} \left(23 K^3+243 K^4 \rho_c +9 \sqrt{3} \sqrt{5 K^6+46 K^7 \rho_c +243 K^8 \rho_c^2}\right)^{1/3}}\nonumber\\ +\frac{\left(23 K^3+243 K^4 \rho_c +9 \sqrt{3} \sqrt{5 K^6+46 K^7 \rho_c +243 K^8 \rho_c ^2}\right)^{1/3}}{9\ 2^{2/3} K^2}
\end{eqnarray} 
and
\begin{eqnarray}
\zeta_{3,2}=-\frac{8}{81 K}+\frac{49 K}{81\ 2^{2/3} \left(23 K^3+243 K^4 \rho_c +9 \sqrt{3} \sqrt{5 K^6+46 K^7 \rho_c +243 K^8 \rho_c ^2}\right)^{2/3}}\nonumber\\+\frac{\left(23 K^3+243 K^4 \rho_c +9 \sqrt{3} \sqrt{5 K^6+46 K^7 \rho_c +243 K^8 \rho_c ^2}\right)^{2/3}}{162\ 2^{1/3} K^3}
\end{eqnarray}




\begin{table}[h!]

  \centering
  \caption{\bf Other Quantities For N=2 and N=3}
\label{table-2}

\begin{tabular} {|| L  ||   L  ||  L || @{} >{\kern\tabcolsep}c @{}  @{} >{\kern\tabcolsep}c @{} }
\hline
\hline
\rowcolor{black!0}[0pt][0pt] &  &  \\
{\large  Quantity}  &   {\large \ \ \ \ \ \ \   N=2}   & {\large \ \ \ \ \ \ \ \ \ \ \ \ \ \ \ \ \ \ \ \ \ \ \ \ \ \ \ \ \ \ \ \ N=3 }  \\ 
\rowcolor{black!0}[0pt][0pt] &  &  \\
\hline
\hline
\rowcolor{black!0}[0pt][0pt] &  &  \\
\rowcolor{black!0}[0pt][0pt]{\large $\mathcal{J}_{mass}$ } & {\ \ \ \ \ \ \ \ \large $\frac{2p{\rho_c}(1-\rho_c )}{(2-\rho_c)+\zeta_{2,1}} $} & {\ \ \ \ \ \ \ \ \ \ \ \ \ \ \ \ \ \ \ \ \ \ \ \ \ \ \ \ \ \ \large $\frac{27p\rho_c(1-\rho_c)}{16\zeta_{3,1}+9(3+2\zeta_{3,2}-2\rho_c)}$ }\\
\rowcolor{black!0}[0pt][0pt] &  & \\
\rowcolor{black!0}[0pt][0pt] &  & \\
\rowcolor{black!0}[0pt][0pt]{\large $\langle\ell\rangle$ } & { \ \ \ \ \ \ \ \ \ \large $\frac{2\rho_c}{{\rho_c + \zeta_{2,1}}}$ } & {\ \ \ \ \ \ \ \ \ \ \ \ \ \ \ \ \ \ \ \ \ \ \ \ \ \ \ \ \ \ \ \ \large $\frac{27 \rho_c}{(9 \rho_c + 16 \zeta_{3,1} + 9 \zeta_{3,2}}$ }\\
\rowcolor{black!0}[0pt][0pt] &  &  \\
\rowcolor{black!0}[0pt][0pt] &  &  \\
\rowcolor{black!0}[0pt][0pt] {\large $\sigma_{\ell}$ {\small (standard deviation)} } & {\ \ \ \large $\sqrt{2} \sqrt{-\frac{{\zeta_{2,1}} ({\zeta_{2,1}}-\rho )}{({\zeta_{2,1}}+\rho )^2}}$ } & {\ \ \ \ \large $\sqrt{3} \sqrt{\frac{-224 {\zeta_{3,1}}^2+81 {\zeta_{3,2}} (-2 {\zeta_{3,2}}+\rho_c )+{\zeta_{3,1}} (-414 {\zeta_{3,2}}+306 \rho_c )}{(16 {\zeta_{3,1}}+9 ({\zeta_{3,2}}+\rho_c ))^2}}$}\\
\rowcolor{black!0}[0pt][0pt] &  &  \\
\rowcolor{black!0}[0pt][0pt] &  &  \\
\rowcolor{black!0}[0pt][0pt]{\large  $ \mathcal{R}$} & {\large $\frac{\sqrt{\frac{{\zeta_{2,1}} (-{\zeta_{2,1}}+\rho_c )}{({\zeta_{2,1}}+\rho_c )^2}} ({\zeta_{2,1}}+\rho_c )}{\sqrt{2} \rho_c }$ } & {\Large $\frac{(16 {\zeta_{3,1}}+9 ({\zeta_{3,2}}+\rho_c )) \sqrt{\frac{-224 {\zeta_{3,1}}^2+81 {\zeta_{3,2}} (-2 {\zeta_{3,2}}+\rho_c )+{\zeta _{3,1}} (-414 {\zeta_{3,2}}+306 \rho_c )}{(16 {\zeta_{3,1}}+9 ({\zeta_{3,2}}+\rho_c ))^2}}}{9 \sqrt{3} \rho_c }$}\\
\rowcolor{black!0}[0pt][0pt] &  & \\
\hline
\hline
\end{tabular}

\end{table}
\end{widetext}

The specific flux profiles under PBC for a few different values of the stickiness parameter $K$ are plotted in 
Fig.\ref{fig2} against the coverage density $\rho_{c}$ for N=2 and N=3. The analytical predictions made by 
the MFT are in excellent agreement with the corresponding MC data obtained for the same set of numerical 
values of the model parameters.
For conventional single-species TASEP for hard rods of length $\ell$, under PBC, the maximum of $\mathcal{J}_{mass}-\rho_c$ curve appears at the coverage density $\rho_c^{*}=\sqrt{\ell}/(\sqrt{\ell}+1)$.   
 Under PBC, in the limit $K \to 0$, rods longer than ${\ell}=1$ are practically non-existent in the NESS, irrespective of the initial conditions, and hence $\rho_c^* \to 1/2$.  But in the opposite limit $K \to \infty$, $\rho_c^* \to \sqrt{2}/(\sqrt{2}+1)$ and $\rho_c^*\to \sqrt{3}/(\sqrt{3}+1)$  for N=2 and N=3, respectively. Thus, by tuning $K$, we can induce a transition from a regime where the track is populated almost exclusively by particles (i.e., rods of length ${\ell}=1$) to a regime where practically all the rods on the lattice have length ${\ell}=N$. Hence, over the range $0.1<K<10$ regime, where $f_u$ and $f_i$ are comparable, a heterogeneous dynamic population of all species of rods having lengths ${\ell}=1,2,...N$ is observed (see Fig.{\ref{fig3}}).

Since it is too difficult to carry out the analytical calculations in the $N \to \infty$ limit, i.e, when the rods can grow by fusion without constraint on the length, we have obtained the results in this limit only by MC simulations. Note that the coverage density $\rho_c$ is conserved by the dynamics of the system under PBC. Therefore, under the PBC with track length L and coverage density $\rho_c$, the maximum length to which a rod can grow is $\rho_c  L$ (which corresponds to a single rod formed by the fusion of all the particles). As $K$ increases, the mean size $<{\ell}>$ of the rods increases (see Fig.\ref{nfig1}). In the limit $K \gg 1$, almost all the rods do merge to form a single rod of length $\simeq \rho_c  L$.  In Fig.\ref{nfig1}(c), which corresponds to $K=500.0$ and $L=1000$, the most probable length of the rod for $\rho_c=0.1$ is, indeed, $\rho_c L = 100$, whereas the probability of finding a rod of any other length is practically vanishingly small. Thus, under PBC, apart from K, another factor that governs the length distribution is the coverage density $\rho_c$.

\begin{widetext}

\begin{figure}[ht]
\begin{center}
\includegraphics[width=1.0\columnwidth]{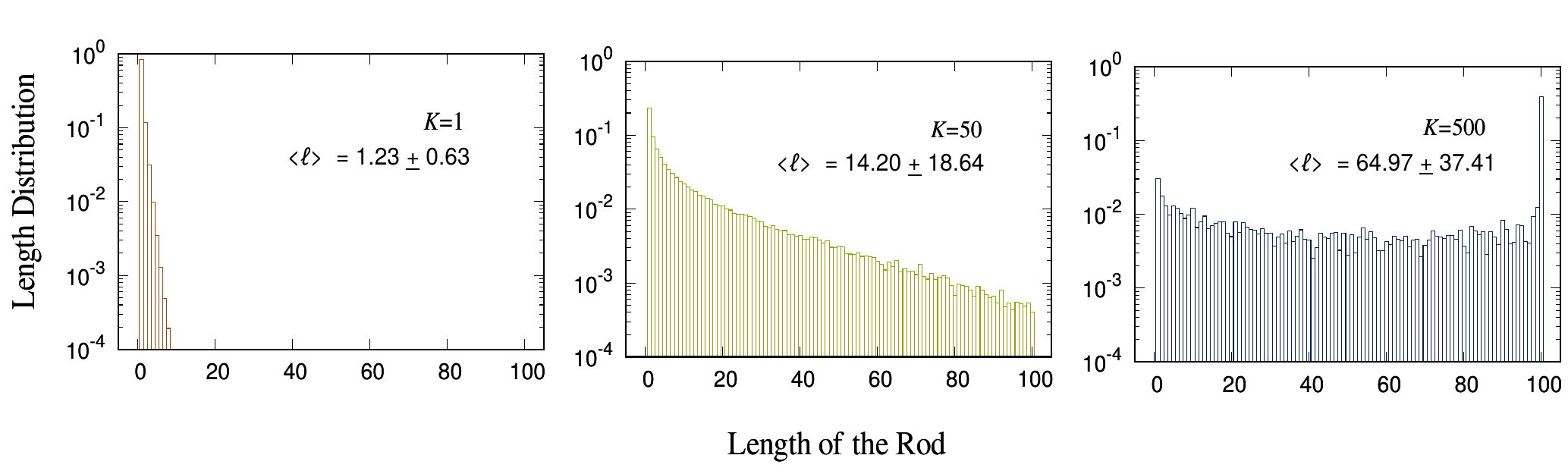}
\end{center}
\caption{(Color online) Steady-state distribution of the lengths of the rods in limit $N \to \infty$ under PBC with (a) $K=1$ (b) $K=50$ (c) $K=500$. For all the three cases, $\rho_c=0.1$ and $L=1000$. }
\label{nfig1}
\end{figure}

\begin{figure}[ht]
\begin{center}
\includegraphics[width=0.5\columnwidth]{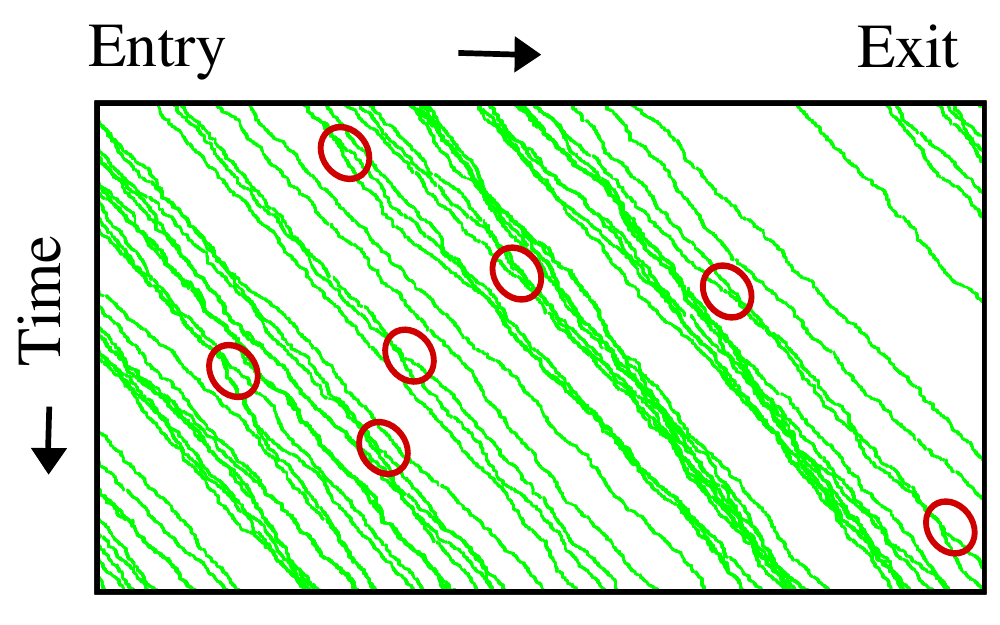}
\end{center}
\caption{(Color online)  {\bf Space-Time Diagram:} Each line depicts the position of a rod, observed as function of time, in the MC simulation of the model with N=3. The merging and splitting of the lines, marked by the red circles, are direct visual evidence of fusion and fission, respectively, of the rods. Parameters used are $L=1000$, 
$\alpha=0.01$, $\beta=0.5$, $f_u=f_i=0.05$, $p$=0.5, MC steps=1000}
\label{fig4}
\end{figure}

\section{Results under OBC} 

The master equations for the model with OBC, under MFA, in the special case $N=2$ are as follows:

For $i=1$ :

\begin{equation}
\frac{P_1(i,t)}{dt}=\alpha (1-P_1(i)-P_2(i))-p P_1(i) \xi-fu P_1(i) P_1(i+1)+f_i P_2(i) 
\label{eq-OBC1}
\end{equation}

\begin{equation}
\frac{P_2(i,t)}{dt}=fu P_1(i) P_1(i+1)-pP_2(i) \xi -f_i P_2(i)
\end{equation}

For $i=2$ to $i=L-2$ :

\begin{eqnarray}
\frac{P_1(i,t)}{dt}=p P_1(i-1)\xi -p P_1(i)\xi-(fu P_1(i) P_2(i+1))\nonumber\\
-(fu P_1(i) P_1(i-1)))+fi P_2(i)+fi P_2(i-1)
\end{eqnarray}

\begin{equation}
 \frac{P_2(i)}{dt}=p P_2(i-1)\xi -p P_2(i) \xi +(fu P_1(i) P_1(i+1))-fi P_2(i)
\end{equation}

For $i=L-1$  :

\begin{eqnarray}
\frac{P_1(i,t)}{dt}=pP_1(i-1)\xi -p P_1(i+1)(1-P_1(i+1)-P_2(i+1))-fu P_1(i) P_1(i-1)+f_i P_2(i)+f_i P_2(i-1)
\end{eqnarray}

\begin{equation}
\frac{dP_2(i)}{dt}=pP_2(i-1)\xi-pP_2(i)-f_iP_2(i)
\end{equation}

For $i=L$ :

\begin{equation}
\frac{dP_1(i)}{dt}=pP_1(i-1)(1-P_1(i)-P_2(i))-\beta P_1(i)+f_iP_2(i-1)
\end{equation}

\begin{equation}
\frac{dP_2(i)}{dt}=pP_2(i-1)-\beta P_2(i)
\label{eq-OBClast}
\end{equation}

The master equations for the model with OBC, under MFA, in the special case $N=3$ are as follows:

For $i=1$ :

\begin{eqnarray}
&\frac{dP_1(i)}{dt}=\alpha(1-P_1(i)-P_2(i)-P_3(i))-pP_1(i)\xi -f_uP_1(i)P_2(i+1) +f_iP_2(i)\nonumber\\\nonumber\\
&\frac{dP_2(i)}{dt}=-pP_2(i)\xi-f_iP_2(i)+f_uP_1(i)P_1(i+1)+\frac{f_i}{2}P_3(i)-f_uP_2(i)P_1(i+2)\nonumber\\\nonumber\\
&\frac{dP_3(i)}{dt}=-pP_3(i)\xi +f_uP_2(i)P1(i+2)+f_uP_1(i)P_2(i+1)-f_iP_3(i)
\end{eqnarray}

For $i=2$ :

\begin{eqnarray}
&\frac{dP1(i)}{dt}=pP_1(i-1)\xi-pP_1(i)\xi -f_uP_1(i)P_1(i+1)-f_uP_1(i)P_1(i-1)\nonumber\\
&-f_uP_1(i)P_2(i+1)+f_iP_2(i)+f_iP_2(i-1)+\frac{f_i}{2}P3(i)\nonumber\\\nonumber\\
&\frac{dP_2(i)}{dt}=(pP2(i-1)\xi-pP_2(i)\xi)+f_uP_1(i)P_1(i+1)+\frac{f_i}{2}(P_3(i)+P_3(i-1))-\frac{f_i}{2}P_2(i)\nonumber\\
&-f_uP_2(i)P_1(i+2)-f_uP_2(i)P_1(i-1)\nonumber\\\nonumber\\
&\frac{dP3(i)}{dt}=(pP_3(i-1)\xi-pP_3(i)\xi)+fuP_2(i)P1(i+2)+f_uP_1(i)P_2(i+1)\nonumber\\
&-f_iP_3(i)
\end{eqnarray}

For $i=3$ to $i=L-3$ :

\begin{eqnarray}
&\frac{dP_1(i)}{dt}=pP_1(i-1)\xi -pP_1(i)\xi -f_uP_1(i)P_1(i+1)-f_uP_1(i)P_1(i-1)\nonumber\\
&-f_uP_1(i)P_2(i+1)-f_uP_1(i)P_2(i-2)+f_iP_2(i)\nonumber\\&+f_iP_2(i-1)+\frac{f_i}{3}P_3(i)+\frac{f_i}{2}P_3(i-2)\nonumber\\\nonumber\\
&\frac{dP_2(i)}{dt}=(pP_2(i-1)\xi- pP_2(i)\xi)+f_uP_1(i)P_1(i+1)+\frac{f_i}{2}(P_3(i)+P_3(i-1))-f_iP_2(i)\nonumber\\
&-f_uP_2(i)P_1(i+2)-f_uP_2(i)P_1(i-1)\nonumber\\\nonumber\\
&\frac{dP_3(i)}{dt}=pP_3(i-1)\xi -pP_3(i)\xi +f_uP_2(i)P_1(i+2)+f_uP_1(i)P_2(i+1)\nonumber\\
&-f_iP_3(i) 
\end{eqnarray}

For $i=L-2$ :

\begin{eqnarray}
&\frac{dP_1(i)}{dt}=pP_1(i-1)\xi- pP_1(i)(1-P_1(i+1)-P_2(i+1)-P_3(i+1))-f_uP_1(i)P_1(i+1)\nonumber\\
&-f_uP_1(i)P_1(i-1)-f_uP_1(i)P_2(i+1)-f_uP_1(i)P_2(i-2)\nonumber\\
&+f_iP_2(i)+f_iP_2(i-1)+\frac{f_i}{2}P_3(i)+\frac{f_i}{2}P_3(i-2)\nonumber\\\nonumber\\
&\frac{dP_2(i)}{dt}=pP_2(i-1)\xi - pP_2(i)(1-P_1(i+1)-P_2(i+1)-P_3(i+1))+f_uP_1(i)P_1(i+1)\nonumber\\
&+\frac{f_i}{2}(P_3(i)+P_3(i-1))-\frac{f_i}{2}P_2(i)-f_uP_2(i)P_1(i-1)\nonumber\\\nonumber\\
&\frac{dP_3(i)}{dt}=pP_3(i-1)\xi -pP_3(i)+f_uP_2(i)P1(i+2)+f_uP_1(i)P_2(i+1)\nonumber\\
&-f_iP_3(i)\nonumber\\
\end{eqnarray}

For $i=L-1$ :

\begin{eqnarray}
&\frac{dP_1(i)}{dt}=pP_1(i-1)(1-P_1(i)-P_2(i)-P_3(i))-pP_1(i)(1-P_1(i+1)-P_2(i+1)-P_3(i+1))-f_uP_1(i)P_1(i-1)\nonumber\\
&-f_uP_1(i)P_2(i-2)+f_iP_2(i)+f_iP_2(i-1)+\frac{f_i}{2}P_3(i)+\frac{f_i}{2}P_3(i-2)\nonumber\\\nonumber\\
&\frac{dP_2(i)}{dt}=pP_2(i-1)(1-P_1(i)-P_2(i)-P_3(i))-pP_2(i)(1-P_1(i+1)-P_2(i+1)-P_3(i+1))\nonumber\\
&+\frac{f_i}{2}(P_3(i)+P_3(i-1))-f_iP_2(i)-f_uP_2(i)P_1(i-1)\nonumber\\\nonumber\\
&\frac{dP_3(i)}{dt}=pP_3(i-1)-pP_3(i)-f_iP_3(i)
\end{eqnarray}

For $i=L$ :
\begin{eqnarray}
&\frac{dP_1(i)}{dt}=pP_1(i-1)(1-P_1(i)-P_2(i)-P_3(i))-\beta P_1(i)+f_iP_2(i-1)+\frac{f_i}{2}P_3(i-2)\nonumber\\\nonumber\\
&\frac{dP_2(i)}{dt}=pP_2(i-1)-\beta P_2(i)+\frac{f_i}{2}P_3(i-1)\nonumber\\\nonumber\\
&\frac{dP_3(i)}{dt}=pP_3(i-1)-\beta P_3(i)
\end{eqnarray}

\end{widetext}

\begin{figure}[ht]
\begin{center}
\includegraphics[width=0.75\columnwidth]{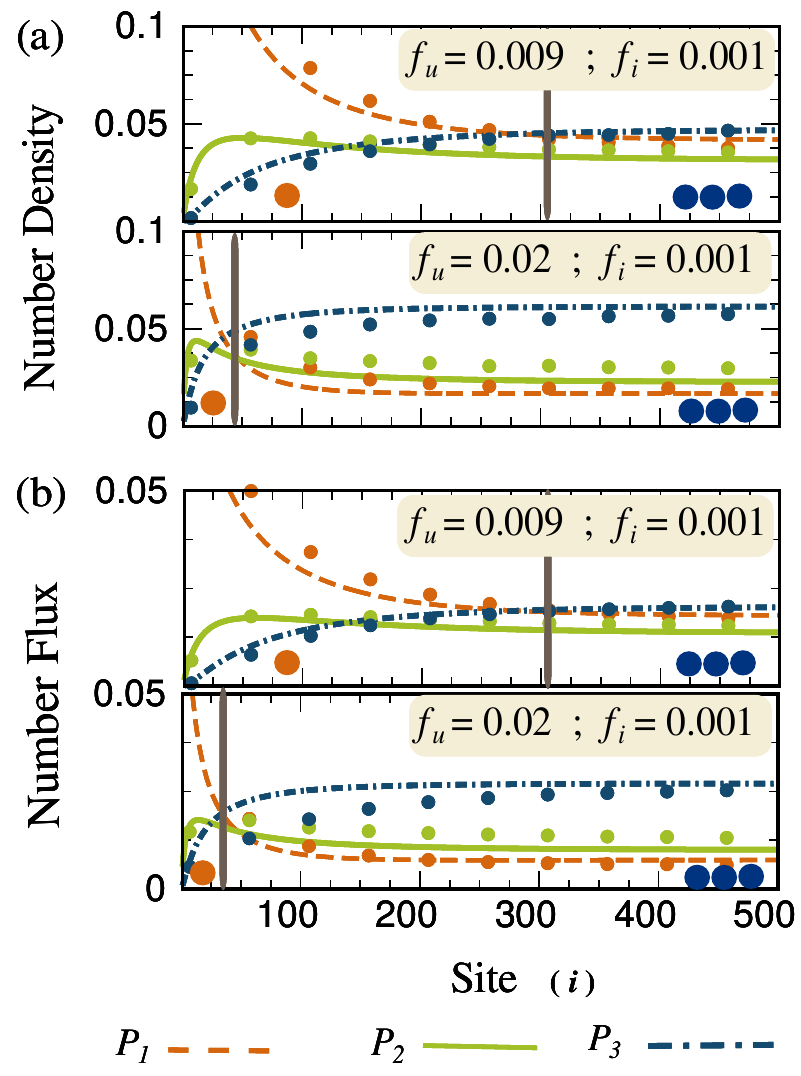}
\end{center}
\caption{(Color online) (a) Number-Density Profile, and (b) Number-Flux Profile under OBC for $N=3$ and  different combination of $f_u$ and $f_i$ mentioned in the figures. $\alpha=0.15$ and $\beta=0.85$ for all the cases. Here we are showing the first 500 sites of system of total length $L$=1000. 
(Lines-MFT ; Dots-Simulation). }
\label{fig5} 
\end{figure}


\begin{figure}[h]
\begin{center}
\includegraphics[width=0.9\columnwidth]{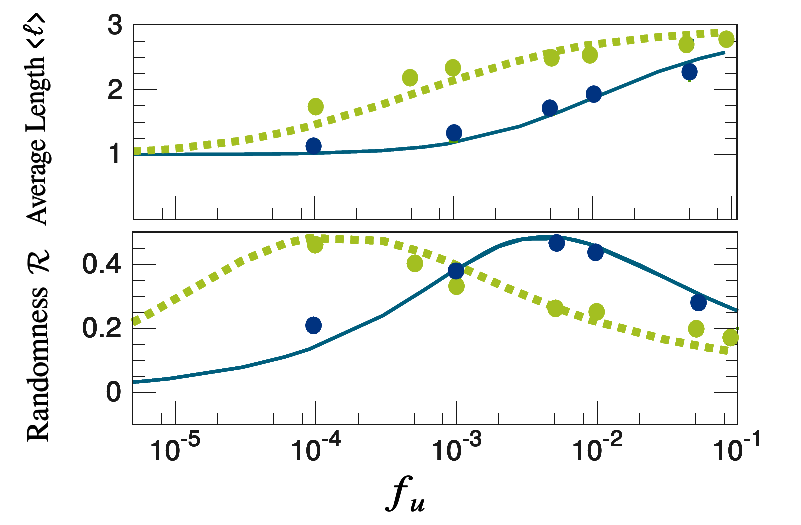}
\end{center}
\caption{(Color online) Mean field predictions for (a) average length $<{\ell}>$ of rods, and (b) randomness parameter $\mathcal{R}$ associated with rod length are plotted against fusion rate $f_u$ for fixed values $N=3$ and $f_i=10^{-4}$. $\alpha=0.9,\beta=0.05$ correspond to $\rho_c \simeq 0.9$ (dotted green ) whereas $\alpha=0.05,\beta=0.5$ correspond to $\rho_c \simeq 0.1$ (continuous blue). (Lines-MFT ; Dots-Simulation). }
\label{fig6}
\end{figure}

The trajectories of the rods, shown in the form of space-time plots in Fig.\ref{fig4}, clearly show fusion and fission events.
Unlike the results obtained under PBC, which were dependent on $f_u$ and $f_i$ only through the ratio $K=f_u/f_i$, those under OBC depend on the individual rates $f_u$ and $f_i$. 
By fixing $\alpha=0.15$ and $\beta=0.85$, we have obtained the number-density profile and number flux profile plotted in Fig.{\ref{fig5}}. In the fusion dominated regime, as the rods of length ${\ell}=1$ enter through $i=1$ and move forward, the population of longer rods increase at the expense of that of smaller rods because of the dominance of fusion. The populations of all the $N$ species continue to evolve over a Transition Zone (TZ) at the end of which, marked by a vertical line in  Fig.{\ref{fig5}}, they attain their respective stationary values.  Rods of length $N$ dominate the population beyond the TZ if the coverage density is high enough for facilitating fusion. The dependence of the width of the TZ on the parameter $N$ and on the rates of the various kinetic processes are discussed in detail in the next section.
 
Beyond the TZ, where the populations of the rods of different length remain stationary, we computed the average length of the rods $<{\ell}>$ as well as the randomness parameter $\mathcal{R}$ (See Fig.{\ref{fig6}}). By fixing the magnitude of $f_i$, we varied $f_u$ over several orders of magnitudes. The average length of the rods varied from $<{\ell}>=1.00$ in the fission dominated regime to $<{\ell}>=3.00$ in the fusion dominated regime (maximum allowed length being $N=3$). But the randomness parameter $\mathcal{R}$ exhibits a nonmonotonic variation with increasing $f_u$. Since fusion of two rods is allowed only if they touch each other, a higher coverage density $\rho_c$ is expected to facilitate more frequent fusion. This is consistent with Fig.{\ref{fig6}}(a) where, for a given $f_{u}$, $<{\ell}>$ is higher at higher $\rho_c$. Moreover, in Fig.{\ref{fig6}}(b) the curve essentially shifts laterally rightward because for attaining the same value of  $\mathcal{R}$ at lower coverage density a higher rate of fusion is required.

\section{Transition Zone}

Entry of monodisperse particles, and their transformation into polydisperse population, through fusion-fission kinetics, gives rise to a special region near the entry site that we have referred to as `transition zone' (TZ). In the fusion dominated regime, as the rods of length ${\ell}=1$ enter through $i=1$ and move forward, the population of longer rods increase at the expense of that of smaller rods because of the dominance of fusion. The populations of all the $N$ species continue to evolve over the TZ, beyond which they attain their respective stationary values.

\subsection{N-dependence of the Transition Zone}

\begin{figure*}
\begin{center}
\includegraphics[width=1.8\columnwidth]{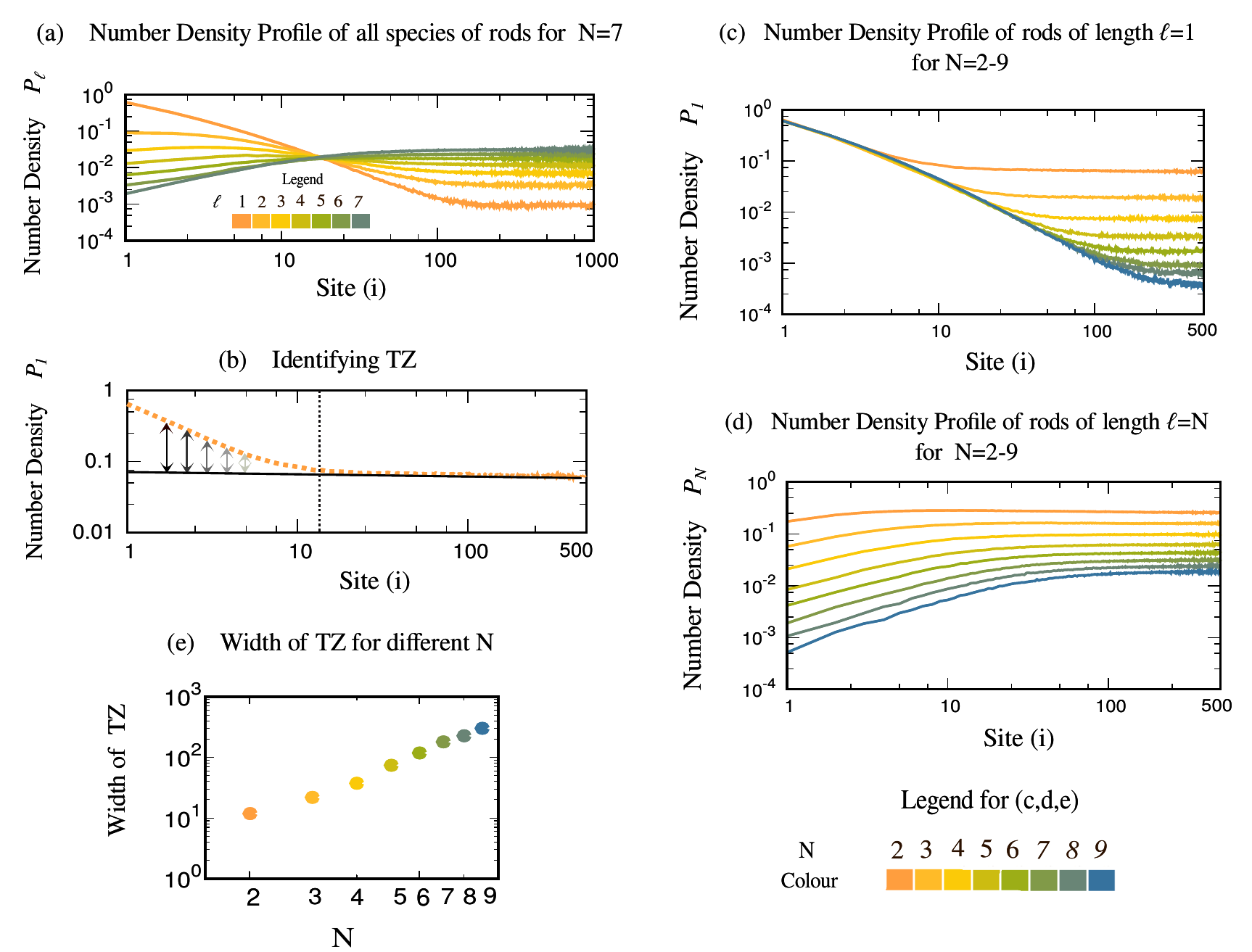}
\end{center}
\caption{(Color online) {\bf Transition Zone (TZ):} (a) Number density profiles of all species of rods of lengths ${\ell} = 1,2,\dots N$ for $N$=7. (b) The dotted vertical line on the number density profile of the particles (i.e., rods of length $\ell =1$) identifies the right boundary of the TZ. (c) Number density profiles of the shortest rods (of length ${\ell}=1$) for $N=2-9$. (d) Number density profiles of the longest rods (of length $\ell=N$) for  $N=2-9$.  (e) Width of TZ for  $N=2-9$. For all the figs.(a)-(e), 
$\alpha=0.9$, $\beta=0.5$, $p=0.5$, $f_u=0.1$ and $f_i=0.0001$.}
\label{nfig2}
\end{figure*}

Since analytical treatment becomes increasingly difficult with increasing value of $N$, we have investigated the $N$-dependence of the width of the TZ only by MC simulation; the result is shown in Fig.\ref{nfig2}. In Fig.{\ref{nfig2}(a)}, the number density profiles for the rods of length ${\ell}=1,2,\dots 7$ are shown on a log-log plot for $N=7$. It is clearly visible on Fig.{\ref{nfig2}}(a) that the last species to achieve stationary density is the  particle (i.e., rod of $\ell=1$). By the distance from the entrance where the density of the particles achieve stationary value, that of all the other longer rods have already achieved their stationary values. Therefore, for quantitative purposes, we identify the right boundary of the TZ as the location where the density of the particles (i.e., rods of $\ell=1$) attain its  stationary value (see Fig.{\ref{nfig2}}(b)). 
In Fig.{\ref{nfig2}}(c) we have drawn the number density profile of the shortest rods (i.e.,${\ell}=1$) and in Fig.{\ref{nfig2}}(d) that of the longest rods (i.e., ${\ell}=N$), both for  $N=2-9$. The variation of width of the TZ with $N$ is shown on a log-log plot in Fig.{\ref{nfig2}}(e). The monotonic increase of the width with $N$ arises from the fact that the broader the possible polydispersity of the rod size, the longer it takes to reach the steady distribution after the entry of the rods as a  monodisperse population. \\

\subsection{Fusion-dependence of the Transition Zone}

To get an intuitive understanding of the dependence of the TZ on the kinetic parameters of the model, we first derive  approximate equations for $P_{\mu}(x)$ ($\mu=1,2$), for the simplest case of $N=2$, in the fusion-dominated regime ($f_{u} \gg f_{i}$) where, in addition, $f_{i}$ is negligibly small. The master equation for $P_{1}(x,t)$ in discrete time and and discretized space is given by 
\begin{widetext}
\begin{equation}
P_1(x,t)=[p\{1-\rho(x)\}\Delta t]P_1(x-\Delta x,t-\Delta t)+[1-p\{1-\rho(x+\Delta x)\}\Delta t-f_u\Delta t]P_1(x,t)
\label{P1_me}
\end{equation} 
\end{widetext}
where $\Delta t$ is the duration of each time step and $\Delta x = \lambda$ is the separation between the successive points in the discretized one-dimensional space. The first term on the right hand side of Eq.(\ref{P1_me}) leads to the gain of $P_{1}$ at $x$ at time $t$ due to the incoming particles of length ${\ell}=1$ that were located at $x-\Delta x$ at time $t-\Delta t$. The second term states that the particle at $x$ neither hopped out rightwards to $x+\Delta x$ nor fused with a neighbouring particle.
Note that the possibility of fusion, captured by the last term, requires that another particle must be adjacent to the particle under consideration with which fusion can take place. However, that factor does not appear explicitly in this term implying that the probability of finding another particle adjacent to the particle of interest has been assumed to be, effectively, unity; this is a reasonably good approximation only at sufficiently high values of $P_{1}(x)$. 
Similarly, the corresponding master equation in discretized space and time is given by  
\begin{widetext}
\begin{equation}
 P_2(x,t) = [p\{1-\rho(x+\Delta x)\}\Delta t]P_2(x-\Delta x,t-\Delta t)+ [1-p\{1-\rho(x+2\Delta x)\}\Delta t]P_2(x,t)+\{f_{u}\Delta t \}\underbrace{P_1(x,t)}_{=\{\rho-2P_2(x,t)\}}
\label{P2_me}
\end{equation} 
\end{widetext}
Note that in writing these equations, the gain of $P_{1}(x)$ and loss of $P_{2}(x)$ that could arise by fission of rods of length ${\ell}=2$ is neglected because, in the fusion-dominated regime of our interest here, $f_i \ll f_u$. Moreover, since N =2, loss of $P_{2}$ by fusion is not possible. Thus, strictly speaking, eqs.(\ref{P1_me})-(\ref{P2_me}) are not applicable to any $N > 2$ even in the fusion-dominated regime.

The equations (\ref{P1_me}) and (\ref{P2_me}) are based on much cruder approximation than those used in 
writing the equations (\ref{eq-OBC1})-(\ref{eq-OBClast}). However, as we show below, the analytical solutions of the equations (\ref{P1_me}) and (\ref{P2_me}) provide more direct intuitive understanding of the variation of the TZ with 
the kinetic parameters of the model than that conveyed by the numerical solutions of (\ref{eq-OBC1})-(\ref{eq-OBClast}). Carrying out a straightforward Taylor series expansion and then imposing the steady state condition 
$\partial P_1(x,t)/\partial t=0=\partial P_2(x,t)/\partial t$, we get the approximate solutions 
\begin{widetext}
\begin{eqnarray}
P_1(x) &=& P_1(0)exp\biggl[{-\frac{f_u}{p\{1-\rho(x)\}+f_{u}}\biggl(\frac{x}{\lambda}\biggr)\biggr]} \nonumber \\
P_2(x) &=& \frac{\rho(x)}{2}+\biggl\{P_2(0)-\frac{\rho(0)}{2}\biggr\}exp\biggl[-\frac{2f_u}{p\{1-\rho(x)\}+2f_u} \biggl(\frac{x}{\lambda}\biggr)\biggr]
\label{Exp-fall_P1_P2}
\end{eqnarray}
\end{widetext}
where $\Delta x = {\lambda}$ is the spacing between successive lattice sites while $P_{1}(0) = P_{1}(x=0)$ and $P_{2}(0) = P_{2}(x=0)$ are the probabilities of finding rods of length ${\ell}=1$ and ${\ell}=2$, respectively, at the left edge of the lattice. The solutions (\ref{Exp-fall_P1_P2}) also ensure that 
\begin{equation}
\rho(x)=P_1(0)+2P_2(0)
\end{equation} 

\begin{figure}[h]
\begin{center}
\includegraphics[width=1.0\columnwidth]{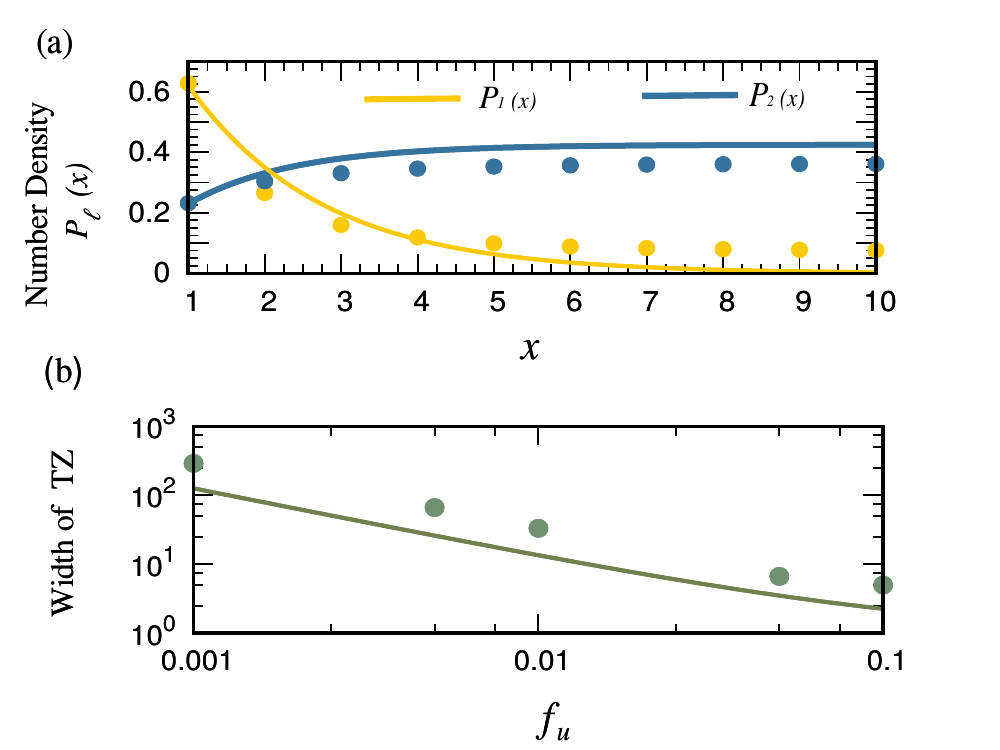}
\end{center}
\caption{(Color online) (a) Number-Density Profiles under OBC in the TZ for $N=2$, obtained from the analytical solutions in equation (\ref{Exp-fall_P1_P2}), are drawn  for $f_u=0.1$, $f_i=0.0001$, $\alpha=0.9$ and $\beta=0.1$. The distance $x$ is expressed in the units of the lattice spacing $\lambda$. Here we are showing the first 10 sites of the system of total length $L$=1000. 
(Lines-Analytical Solutions in equation(\ref{Exp-fall_P1_P2}); Dots-Simulation). (b) Variation of the width of the TZ with $f_{u}$, for the fixed values  $\alpha=0.9$ , $\beta=0.1$ and $f_i=0.0001$, is displayed on a log-log plot to establish the exponenetial decrease of the width with the increasing rate of fusion (Line-Analytical Expression for $TZ_{width}$ in equation (\ref{tz-width}) ; Dots-Simulation).  }
\label{nfig3}
\end{figure}

This derivation is based on a crude assumption i.e, $\rho(x)=\rho(0)$. Moreover, for the subsequent analysis, we treate $P_{1}(0)$ and $P_{2}(0)$ in (\ref{Exp-fall_P1_P2}) as fitting parameters which we fix by using the values of $P_{1}(0)$ and $P_{2}(0)$ available from MC simulations. In the fusion-dominated regime, we get very good agreement between $P_{1}(x)$, $P_{2}(x)$ predicted by (\ref{Exp-fall_P1_P2}) and the corresponding numerical data obtained from MC simulations of the model for $N=2$ (see Fig.\ref{nfig3}(a)) over the entire TZ.

The prescription introduced in the preceeding subsection for identifying the width of TZ and the exponential behaviour of the number density of the rods of length ${\ell}=1$ imply that the width of the TZ for $N=2$  is 
\begin{equation}
TZ_{width}\approx\frac{[p\{1-\rho(0)\}+f_u]{\lambda}}{f_u}
\label{tz-width}
\end{equation}
Approximate linear variation of the width of TZ with the fusion rate $f_{u}$ in the log-log plot of Fig.\ref{nfig3}(b) is consistent with the power-law dependence of the width on $f_{u}$ in eq.(\ref{tz-width}).

\section{Phase Diagram} 

The analytical results obtained under PBC are exploited in plotting the phase diagram under OBC using the {\it extremum current hypothesis} (ECH) \cite{krug91,popkov99,antal00,popkov01,hager01a,hager01b}. For the implementation of ECH, first, one needs the expression for $\rho_{c}^*$ {(Fig.{\ref{fig3}}) }. Next, one imagines a scenario where the entry and exit points of the actual physical system are assumed to be coupled to two mass reservoirs of densities $\rho_-$ and $\rho_+$ respectively (see Fig.{\ref{fig1})}, and calculates $\rho_{\pm}$ using the expressions for flux derived under PBC. Equating both the incoming and outgoing flux at entry site 1, we solve for $\rho_-$ as a function of entry rate $\alpha$, $f_u$ and $f_i$. Similarly, equating the incoming and outgoing flux at exit site $L$ gives $\rho_+$ as function of exit rate $\beta$, $f_u$ and $f_i$.

\subsection{Steps for Plotting the Phase Diagram}
\subsubsection{Expressions for $\rho_+$ and  $\rho_-$}

Let us assume that sufficiently close to the left boundary at $i=1$, the number density can be approximated by 
$\rho_{-}$. Therefore,
at the entry site $i=1$, the mass flux $J(1)_{in}$ moving into this site from the reservoir of density $\rho_{-}$ is 
\begin{equation}
J(1)_{in} = \alpha(1-\rho_-). 
\label{eq-Jinat1} 
\end{equation}
The conditional probability of finding an empty site, provided that $\ell$ sites on its left are covered by a rod of length $\ell$ is $P_{\ell}(\overbrace{\underline{1........1}}^{\ell}|{0})$. Therefore, the mass flux $J(1)_{out}$ moving out of the same site $i=1$ is given by 
\begin{equation}
J(1)_{out}=\rho_{-} \mathcal{P}_{\ell}(\overbrace{\underline{1........1}}^{\ell}|{0})
\label{eq-Joutat1}
\end{equation}
where

\begin{equation}
    \mathcal{P}_{\ell}(\overbrace{\underline{1........1}}^{\ell}|{0})=\frac{1-\sum_{j=1}^{N}\lbrace\sum_{k=1}^{j}P_j(i+k)\rbrace}{1+\sum_{j=1}^{N}P_j(i+j)-\sum_{j=1}^{N}\lbrace\sum_{k=1}^{j}P_j(i+k)\rbrace}
\label{conditional_probability_N1}
\end{equation}

Substituting (\ref{conditional_probability_N1}) into (\ref{eq-Joutat1}) and then equating the outgoing flux  (\ref{eq-Joutat1}) with in incoming flux (\ref{eq-Jinat1}) at site 1 ($J(1)_{in}=J(1)_{out}$), we get an equation for $\rho_-$ as a function of $f_u$, $f_i$ and $\alpha$ 

\begin{equation}
\rho_-=\rho_-(\alpha,f_u,f_i)
\label{rho_minus}
\end{equation}

Similarly, $P_{\ell}(\underbrace{1.......1}_{\ell}|\underline{0})$ denotes the conditional probability that, given an uncovered site, there will be  $\ell$ adjacent sites to the left which are covered simultaneously by a rod of length $\ell$. 
We also assume that sufficiently close to the right boundary at $i=L$, the number density can be approximated by 
$\rho_{+}$. 
So incoming flux at site $i=L$ is
\begin{equation}
J(L)_{in}=\rho_{+} \mathcal{P}_{\ell}(\underbrace{1.......1}_{\ell}|\underline{0})
\label{eq-JinatL}
\end{equation}
where
\begin{equation}
    \mathcal{P}_{\ell}(\underbrace{1.......1}_{\ell}|\underline{0})=\frac{\sum_{j=1}^{N}P_j(i+j)}{1+\sum_{j=1}^{N}P_j(i+j)-\sum_{j=1}^{N}\lbrace\sum_{k=1}^{j}P_j(i+k)\rbrace}
\label{conditional_probability_N2}
\end{equation}
and the outgoing flux $J(L)_{out}$ is given by 
\begin{equation}
J(L)_{out}=\beta\rho_{+}.
\label{eq-JoutatL}
\end{equation}

Hence equating the incoming and outgoing fluxes (\ref{eq-JinatL}) and (\ref{eq-JoutatL}), respectively, at $i=L$ and solving the resulting equation for $\rho_{+}$, we get $\rho_{+}$ as a function of $f_u$,$f_i$ and $\beta$
\begin{equation}
\rho_+=\rho_+(\beta,f_u,f_i)
\label{rho_plus}
\end{equation}

\subsubsection{Surface separating LD-MC phases}

According to the ECH, at the boundary between the LD and MC phases 
\begin{equation}
\rho_-(\alpha,f_u,f_i)=\rho_{c}^*
\label{alpha_star_cond}
\end{equation}
Substituting the expressions ({\ref{rho_minus}}) and appropriate $\rho_c^*(f_u,f_i)$ in ({\ref{alpha_star_cond}}) we get the equation
\begin{equation}
\alpha^*=\alpha^*(\rho_c^*,f_u,f_i)
\label{alpha_star}
\end{equation}
for the surface separating the LD and MC phases.

\subsubsection{Surface separating HD-MC phases}

According to the ECH, at the boundary between the HD and MC phases 
\begin{equation}
\rho_+(\beta,f_u,f_i)=\rho_{c}^*
\label{beta_star_cond}
\end{equation}
Substituting the expressions ({\ref{rho_plus}}) and $\rho_c^*(f_u,f_i)$  in ({\ref{beta_star_cond}}) we get the equation
\begin{equation}
\beta^*=\beta^*(\rho_c^*,f_u,f_i)
\label{beta_star}
\end{equation}

for the surface separating the HD and MC phases.

\subsubsection{Surface separating LD-HD phases}
The equation for this surface is obtained by exploiting the fact that exactly on this surface the LD and HD phases coexist simultaneously in the system supporting a single steady state flux that flows through the system. Therefore, the corresponding condition can be implemented mathematically by 
 \begin{equation}
 J_{PBC}(\rho_-(\alpha,f_u,f_i))= J_{PBC}(\rho_+(\beta,f_u,f_i))
 \end{equation}

\begin{figure*}
\begin{center}
\includegraphics[width=1.5\columnwidth]{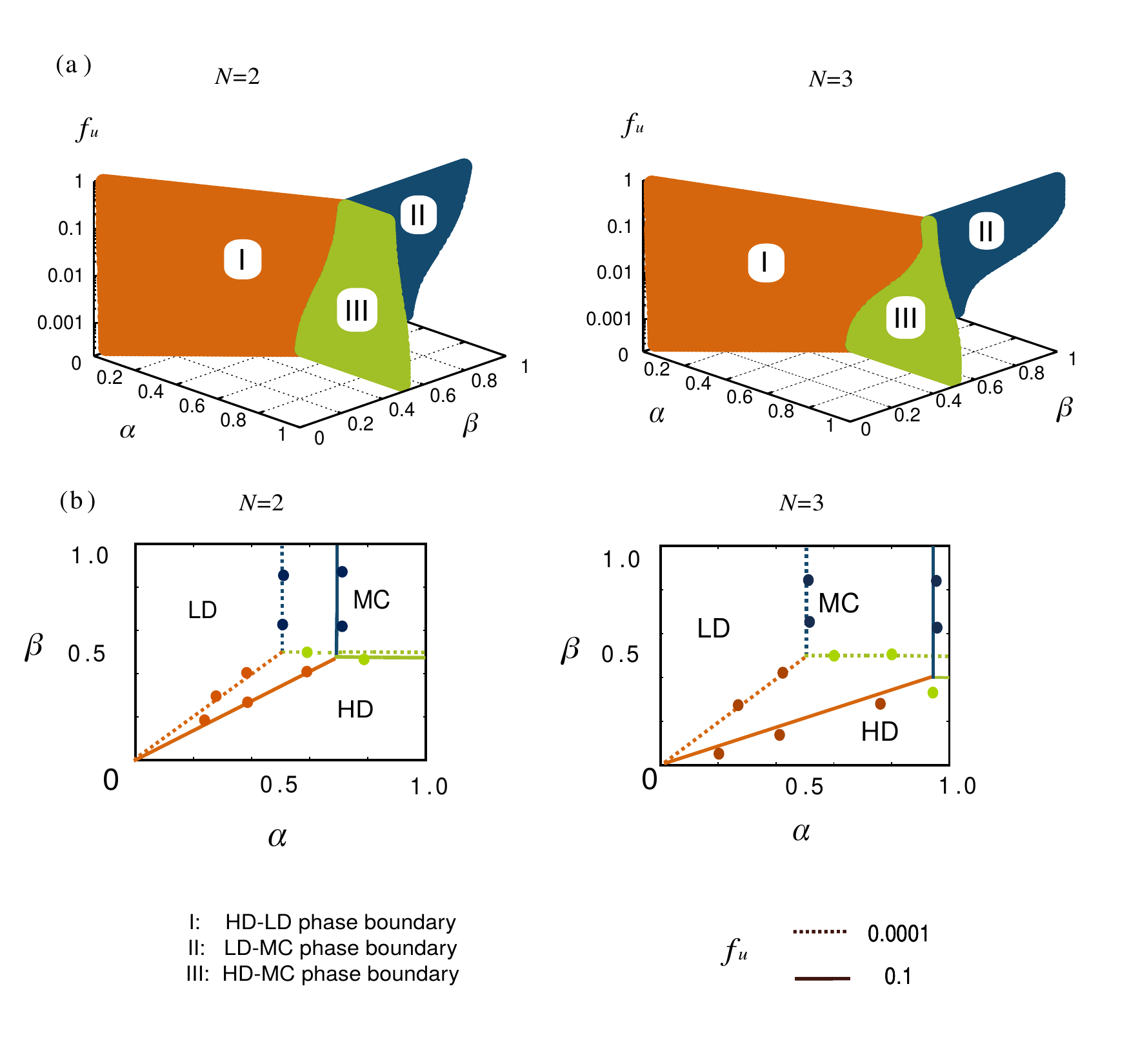}
\end{center}
\caption{(Color online) Phase diagram of the model under OBC for N=2 and N=3 obtained using the extremum current hypothesis. (a) The 3D phase diagram is plotted in $\alpha-\beta-f_{u}$ space.  (b) 2D cross sections of the 3D phase diagram,  for two values of $f_{u}$, are projected on the $\alpha-\beta$ plane. Value of $f_i$ is fixed at 0.01. (Lines-MFT ; Dots-Simulation)}
\label{fig7}
\end{figure*}

\begin{widetext}
\subsection{Phase Boundaries for  N=2 }
Following the above mentioned steps, here we obtained the phase boundaries for our model with N=2. 
As $\alpha^*$ and $\beta^*$ represent the surface separating LD-MC and HD-MC phases respectively, their expressions are as follows:

\begin{equation}
\alpha^*=-\frac{8{f_u} \rho_c^*}{{f_i}-8{f_u}+4 {f_u} \rho_c^* -\sqrt{{f_i}} \sqrt{{f_i}+8 {f_u}\rho_c^* }}
\label{alpha_star_2}
\end{equation}

\begin{equation}
\beta^*=\frac{{f_i}-4 {f_u} \rho_c^* -\sqrt{{f_i}} \sqrt{{f_i}+8 {f_u} \rho_c^* }}{{f_i}-8 {f_u}+4 {f_u} \rho_c^* -\sqrt{{f_i}} \sqrt{{f_i}+8 {f_u} \rho_c^* }}
\label{beta_star_2}
\end{equation}

Whereas the equation for the surface separating the HD-LD phase is as follows: 

\begin{equation}
    \frac{-\alpha  (f_i-8 f_u-4 f_u \alpha )+\alpha \sqrt{f_i} \sqrt{f_i+8 f_u \alpha +4 f_u \alpha ^2}}{ f_u \left(8+8 \alpha +2 \alpha ^2\right)}=\frac{f_i(1-\beta) + 4 f_u \beta +4 f_u \beta ^2+\sqrt{f_i}(-1+\beta ) \sqrt{f_i+4 f_u \beta +4 f_u \beta ^2}}{ f_u \left(2+4 \beta +2 \beta ^2\right)}
\label{alpha-beta_2}
\end{equation}
\end{widetext}

\subsection{Interpreting Phase Diagram}

For conventional $\ell-$TASEP, i.e, TASEP with hard-rods of length $\ell$ only ,the known results are tabulated in Table {\ref{table-3}}.

\begin{table}[h!]
  \centering
  \caption{\bf Results for conventional $\ell$-TASEP}
  \label{table-3}
 \begin{tabular}{c c c c} 
 \hline
 \hline
 $\ell$ & & $\rho_c^*$  & {\small $\alpha^*=\beta^*$ } \\ [0.5ex] 
 \hline
 \hline
 \\
 $\ell$ & & $\frac{\sqrt{\ell}}{\sqrt{\ell}+1}$  & $\frac{1}{\sqrt{\ell}+1}$ \\
 \\
 1 & & 0.5  & 0.5 \\
 2 & & 0.58 & 0.41 \\
 3 & & 0.63 & 0.36 \\
 
 \hline
\end{tabular}
\end{table}

Now to interpret the phase boundaries of our model with $N=2$ in extreme limit, we re-express equations (\ref{alpha_star_2}) and (\ref{beta_star_2}) in terms of $\zeta_{2,1}(\rho_c^*,f_u,f_i)$ and $\rho_c^*$ in Table {\ref{table-4}} and give the approximations in extreme limits of $f_u$ .

\begin{table}[h!]
  \centering
  \caption{\bf Phase Boundaries in Extreme limits of $f_u$ for N=2 case }
  \label{table-4}
 \begin{tabular}{c c c c c c c} 
 \hline
 Quantity & & Expression &  & $f_u \to 0$ &  & $f_u \to \infty$ \\ [0.5ex] 
 \hline
 \\
 $\zeta_{2,1}(\rho_c^*,f_u,f_i)$ & & $\frac{\sqrt{1+8 (f_u/f_i) \rho_c^*}-1}{4(f_u/f_i)}$ & & $\rho_c^*$ & & $\approx 0$\\ 
 \\
 $\alpha^*$ & & $\frac{2\rho_c^*}{2-\rho_c^*+\zeta_{2,1}}$ & & $\rho_c^*$=0.5 & & 2($\frac{\rho_c^*}{2-\rho_c^*})\approx$0.82 \\
 \\
 $\beta^*$ & & $\frac{\rho_c^*+\zeta_{2,1}}{2-\rho_c^*+\zeta_{2,1}}$ & & $\rho_c^*$=0.5 & &$\frac{\rho_c^*}{2-\rho_c^*}\approx$0.41 \\
 
 \hline
\end{tabular}
\end{table}

\begin{table}[h!]
  \centering
  \caption{\bf Phase Boundaries for arbitrary N in $f_u \to \infty $  limit}
  \label{table-5}
 \begin{tabular}{c c c c} 
 \hline
 \hline
 $N$ & & $\alpha^*$  & {\small $\beta^*$ } \\ [0.5ex] 
 \hline
 \hline
 \\
 1 & & 0.5  & 0.5 \\
 2 & & 0.82 & 0.41 \\
 3 & & 1.08 & 0.36 \\
 N & & $\frac{N}{\sqrt{N}+1}$ & $\frac{1}{\sqrt{N}+1}$ \\
 \hline
\end{tabular}
\end{table}
Results in the extreme limits can be understood as follows: In $f_u \to 0$ limit, the phase diagram matches exactly with the phase diagram of particles (rods of $\ell=1$) as the particles upon entry have no tendency to fuse. However, for $f_u > f_i$, all the HD-MC boundary shifts downwards as system tends to, effectively, a single species TASEP with rod-length = N ( N=2,3 in the present case). But LD-MC phase boundary shifts towards right because, in order to form more rods of length $\ell = N$ in the fusion dominated regime, rods of length $\ell = 1$ must enter at a higher rate $\alpha$ so that interparticle distance between them shorten to facilitate effective fusion.
When $f_u \to \infty$ the value of $\beta^*$ matches with that of $N-$TASEP ($\ell$ i.e, N) but $\alpha^*=N\beta^*$ in this limit because for the formation of a rod of $\ell=N$, N particles of $\ell=1$ must enter. Hence, it altogether shifts the LD-MC phase boundary ($\alpha^*$) with increasing $f_u$ in such a way that $\alpha^*>\beta^*$ and becomes $\alpha^* \approx N\beta^*$ in $f_u \to \infty$ limit. The results for our model for arbitrary $N$ are summarised in Table {\ref{table-5}}. Another inference from the phase-diagram is that as we increase $f_u$, it is the region of the phase diagram covered by MC phase shrinks.

Fixing $f_i$=0.01, and taking constant cross sections of this 3D phase diagrams for two different values of $f_u$ we plot the projections of these two cross sections onto the $\alpha-\beta$ plane, as shown in Fig.\ref{fig7}. 
In the limit $f_u << f_i $ the phase boundaries approach those for single-species TASEP with ${\ell}=1$. 
In the opposite limit $f_u > f_i$, all the HD-MC boundary shifts downwards as system tends to, effectively, a single species TASEP with rod-length $\ell=N$ . But LD-MC phase boundary shifts towards right because, in order to form more rods of $\ell=2$ and $\ell=3$ (in case of N=2 and N=3 cases, respectively) in the fusion dominated regime, rods of $\ell=1$ must enter at a higher rate $\alpha$ so that interparticle distance between them shorten to facilitate effective fusion.

\section{Summary, Discussions and Conclusions}
 
Motivated by the fusion and fission of cargoes in intraflagellar transport (IFT), in this paper we have developed a multi-species exclusion model where rods enter the lattice as single particles (i.e., as rods of length ${\ell}=1$), but their length change dynamically because of fusion and fission as the rods hop forward. However, lengths of the rods are not allowed to grow beyond a maximum length $N$. Consequently, in principle, at the exit rods can have lengths ranging from ${\ell}=1$ to ${\ell} = N$ although not with equal probability. 
We have also considered the limit $N \to \infty$ which essentially relaxes the constraint on the maximum rod size. 

Under PBC, we have derived analytical expressions for several quantities that characterize the NESS of the system. 
By a combination of mean-field theory and MC simulations, we have analyzed the density profile and flux profile of the rods in this model under OBC. These results establish the existence of a `transition zone' (TZ) adjacent to the point of entry into the system. The term `transition zone' in our theory should not be confused with the usage of the term in biology to describe a subcellular compartment that is believed to be present between the flagellar base and flagellum.  

Carrying out extensive MC simulations, we demonstrate the dependence of the width of the TZ on the parameter $N$. Moreover, based on a set of approximate analytical arguments, which are well justified for the special case $N=2$ at sufficiently high rate of fusion, we also derive an expression for the width of the TZ. This analytical expression,  and its comparison with MC data, demonstrates how the rates of the kinetic processes control the width of the TZ. 

The agreements between the theoretical predictions  and MC simulations in the NESS of the model are very good, in spite of the mean-field approximations made in writing the master equations on which the theory is based. There are other similar examples of exclusion processes where MFT performs remarkably well; the extreme case being the TASEP under PBC with random-sequential updating for which the mean-field theory turned out to be exact \cite{schad10}. In this work we have focussed exclusively on the NESS of the model which is attained in the sufficiently long time limit. The short-time transient behavior of the model, which has not been studied here, may be important from the perspective of IFT and similar transport phenomena in other types of long cell protrusions. The MFT may require improvements \cite{baker10,markham13}, by incorporating important correlations, to achive sufficiently good agreement with the corresponding MC data.

The Burger's equation \cite{burgers}, which is a nonlinear diffusion equation, is known to provide hydrodynamic description of TASEP \cite{bennaim10,schad10} and related exclusion models \cite{nagel96,schonherr04,schonherr05}. In a future publication, we intend to derive the hydrodynamic counterpart \cite{demasi91} of the discrete `microscopic' exclusion model reported here. Since, at least in principle, the fusion and fission of the rods in our model can be treated as `reactions', the hydrodynamic equations are expected to be some generalized reaction-diffusion equations. Although there are indirect indications in support of this expectation \cite{simpson10,simpson11,penington12}, the actual derivation would be a challenging nontrivial task. 

The model proposed here may be regarded as a model that allows both aggregation and fragmentation \cite{bennaim10} of self-driven clusters. In this terminology, each rod is identified as a ``cluster'' where 
a rod of length ${\ell}=1$ (i.e., a particle) is an ``elemental cluster''. The prescriptions for fusion of the rods in our model corresponds to mass-conserving binary reactions of the clusters. Similarly, the fission of a rod in our model corresponds to binary fragmentation \cite{bennaim10}. Recently, generalizations of TASEP with irreversible aggregation of particles and rods have also been studied \cite{bunzarova17}. The IFT trains are distinct from train-like clusters reported earlier \cite{dong12,pinkoviezky14}. 

However, there are some crucial differences between our model and the widely studied models of aggregation-fragmentation phenomena. The rods are self-driven and, therefore, the system can attain only NESS with non-vanishing flux. Moreover, spatial locations of the rods are very important because only two contiguous rods can fuse. We do calculate the distributions of the lengths of the rods in the NESS, which is the counterpart of cluster-size distribution in aggregation-fragmentation phenomena. However, our attention is also focussed on quantities that are of primary interest in exclusion processes, namely flow properties and density profiles that characterize the phase diagram of the system. So far as the results are concerned, the most important finding of this paper is the existence and nature of the TZ.

Another class of exclusion models with `sticky' particles \cite{garza15,teimouri15} and with sticky rods  \cite{narasimhan17} have been reported in the literature. `Stickiness' in these models arise from the attractive interaction among the particles and rods. In contrast, there is no `attractive interaction' among the particles and rods in our model. Moreover, in refs.\cite{garza15,teimouri15,narasimhan17} particles in a cluster retain their distinct identity and hop independently although the rates of their hop depend whether or not they are part of a cluster before or after the hop. In contrast, particles and rods lose their distinct identify upon fusion with another particle or  rod; the resulting rod emerges with a new identity and hops as a single object.

In our model, at each MCS, the probabilities of forward hopping, fission and exit of a rod as well as the probability of fusion of two rods are all independent of its length. An alternative scenario can be envisaged where, in principle, each of these probabilities can depend on the instantaneous length of the rod(s). Another variant of the model could allow the number of allowed fissions of a rod in each MCS proportional to the length. The current version of the model allows both `severing' anywhere in the bulk and `chipping' from the edges of a rod with equal probability. More restrictive models could allow either `severing' or `chipping'. 

As explicitly stated in the introduction, the model developed here falls short of a complete reslistic description of IFT.
Nevertheless, conceptual and mathematical framework developed here may serve as foundation of the theoretical approach to be adopted for a complete description of IFT in near future \cite{patra18}.

\section*{ACKNOWLEDGEMENT}
SP thanks Soumendu Ghosh and Bhavya Mishra for useful discussions. DC thanks Jonathon Howard for useful suggestions, Gunter Sch\"utz for valuable comments on a preliminary version of this manuscript and Prabal K. Maiti for hospitality at IISc, under the ``DST MathBio'' program, during the preparation of this manuscript. This work is supported by ``Prof. S. Sampath Chair'' Professorship (DC) and a J.C. Bose National Fellowship (DC). 

\section*{REFERENCES}

\clearpage


\appendix
\begin{widetext}
\section{Derivation of Conditional Probability}

The mutual exclusion in our model is captured by $\xi_{N,\ell}(\underline{i}|i+\ell)$ which denotes the conditional probability that the site $i+\ell$ is not covered by another rod, given that site $i$ is occupied by a rod of length $\ell$. 
In the steady state under periodic boundary condition, each site is treated under same footing. In this case, translation invariance follows naturally and $\xi_{N,\ell}(\underline{i}|i+\ell)=\xi_{N,\ell}(\underline{1}|1+\ell)$. So here we present the main steps of our calculation of $\xi_{N,\ell}(\underline{1}|1+\ell)$.

We first consider the special case N=2. Let the symbol $Z(L,N_1,N_2)$ represent the number of ways of arranging $N_1$ rods of length $\ell$=1, $N_2$ rods of length $\ell=$2 and $L-N_1-2N_2$ gaps and it is given by
\begin{equation}
Z(L,N_1,N_2)
=\frac{(N_1+N_2+L-N_1-2N_2)!}{(N_1+N_2)! (L-N_1-2N_2)!}
\end{equation}
Number of ways in which a rod of length $\ell=1$ occupies site $i=1$ is given by $
Z(L-1,N_1-1,N_2)$. Out of these, number of ways in which a rod of length $\ell=1$[$\ell=2$] can occupy site $i=2$ is $Z(L-2,N_1-2,N_2)$ [$Z(L-{1}-{2},N_1-1,N_2-1)$]. Therefore, given that there is a rod of length $\ell=1$ occupying site $i=1$, probability of finding another rod of length $\ell=1$ occupying site $i=2$ is
\begin{equation}
\mathcal{P}(\underline{1}|1+\ell)=\frac{Z(L-2,N_1-2,N_2)}{Z(L-1,N_1-1,N_2)}=\frac{(N_1+N_2-1)}{(L+N_1+N_2-1N_1-2N_2-1)}
\end{equation}
which is also equal to probability of finding a rod of length $\ell=2$ occupying site $i=2$, given that a rod of length $\ell=1$ occupies $i=1$. Therefore, we conclude that probability of finding site $i=2$ uncovered provided site $i=1$ is occupied by a rod of length $\ell=1$ is
\begin{equation}
\xi_{2,1}(\underline{i}|i+1)=\frac{(L-N_1-2N_2)}{(L+N_1+N_2-N_1-2N_2-1)}
\end{equation}

It is straightforward to show that

\begin{equation}
\xi_{2,1}(\underline{i}|i+1)=\xi_{2,2}(\underline{i}|i+2)
\end{equation}
i.e, conditional probability that the site $i+1$ is not covered by another rod, given that there is a rod of length $\ell=1$ occupying site $i$ is equal to the conditional probability that the site $i+2$ is not covered by another rod, given that there is a rod of length $\ell=2$ occupying site $i$.

So, for N=2 using the compact notation $\xi_2$ for $\xi_{2,1}=\xi_{2,2}$, which denotes if a rod of length $\ell$ ($\ell=1/\ell=2$) occupies site $i$, probability of finding site $i+\ell$ uncovered is given by 
\begin{equation}
\xi_{2}(\underline{i}|i+\ell)=\frac{(L-N_1-2N_2)}{(L+N_1+N_2-N_1-2N_2-1)}
\label{cond_prob_2}
\end{equation}

Introducing the number densities $\rho_1$(=$N_1/L$) and $\rho_2$(=$N_2/L$), equation ({\ref{cond_prob_2}}) can be reexpressed as
\begin{equation}
\xi_{2}(\underline{i}|i+\ell)=\frac{1-\rho_1-2\rho_2}{1+\rho_1+\rho_2-\rho_1-2\rho_2}
\label{cond_prob_2_alt}
\end{equation}

Because of fusion and fission, $\rho_1$ and $\rho_2$ keep fluctuating. Therefore, we replace $\rho_1$ and $\rho_2$ with the corresponding occupational probabilities $P_1$ and $P_2$ getting
\begin{equation}
\xi_2(\underline{i}|i+\ell)=\frac{1-\sum_{s=1}^{1}P_1(i+s)-\sum_{s=1}^{2}P_2(i+s)}{1+P_1(i+1)+P_2(i+2)-\sum_{s=1}^{1}P_1(i+s)-\sum_{s=1}^{2}P_2(i+s)}
\label{conditional_probability_2}
\end{equation}

Beauty of the conditional probability (\ref{cond_prob_2_alt}) is that, when $\rho_2=0$, it reduces to  (1-$\rho_1$) which is the conditional probability to be used for exclusion processes with particles i.e, rods of $\ell=1$. Similarly when $\rho_1=0$, the conditional probability (\ref{cond_prob_2_alt}) reduces to $\frac{1-2\rho_2}{1+\rho_2-2\rho_2}$ which is the conditional probability to be used for exclusion processes with hard rods of length $\ell=2$ .

Proceeding similarly, conditional probability for  N=3 is found to be
\begin{equation}
\xi_3(\underline{i}|i+\ell)=\frac{1-\sum_{s=1}^{1}P_1(i+s)-\sum_{s=1}^{2}P_2(i+s)-\sum_{s=1}^{3}P_3(i+s)}{1+P_1(i+1)+P_2(i+2)+P_3(i+3)-\sum_{s=1}^{1}P_1(i+s)-\sum_{s=1}^{2}P_2(i+s)-\sum_{s=1}^{3}P_3(i+s)}
\label{conditional_probability_3}
\end{equation}
Hence, the generalised conditional probability for arbitrary N  is given by
\begin{equation}
    \xi_N(\underline{i}|i+\ell)=\frac{1-\sum_{j=1}^{N}\lbrace\sum_{k=1}^{j}P_j(i+k)\rbrace}{1+\sum_{j=1}^{N}P_j(i+j)-\sum_{j=1}^{N}\lbrace\sum_{k=1}^{j}P_j(i+k)\rbrace}
\label{conditional_probability_N}
\end{equation}

\section{Phase Boundaries for N=3 }

Here we present the phase boundaries for our model with N=3. As $\alpha^*$ and $\beta^*$ represent the surface separating LD-MC and HD-MC phases respectively, their expressions are as follows:

\begin{eqnarray}
&\alpha^*= \frac{A1}{A2+A3+A4+A5}
\label{alpha_star_3}
\end{eqnarray}

\begin{eqnarray}
&\beta^*= \frac{B1+B2+B3+B4}{B1+B2+B3+B5}
\label{beta_star_3}
\end{eqnarray}
where
\begin{equation}
A1=324 {f_i} {f_u}^3 \rho_c^*  \phi^{2/3}
\end{equation}

\begin{equation}
 {A2}=278\ 2^{1/3}  {f_i}^4  {f_u}^4+3\ 2^{2/3} \sqrt{3} \sqrt{ {f_i}^4  {f_u}^6 \left(5  {f_i}^2+46  {f_i}  {f_u} (\rho_c^*)+243  {f_u}^2 (\rho_c^*)^2\right)} {\phi}^{1/3}
\end{equation}

\begin{eqnarray}
 {A3}= {f_i}^2  {f_u}^2 {\phi}^{1/3} \left(81\ 2^{2/3}  {f_u}^2 \rho_c^* -32 {\phi}^{1/3}\right)
\end{eqnarray}

\begin{eqnarray}
 {A4}=12  {f_i} \left(8\ 2^{1/3} \sqrt{3}  {f_u} \sqrt{ {f_i}^4  {f_u}^6 \left(5  {f_i}^2+46  {f_i}  {f_u} \rho_c^* +243  {f_u}^2 (\rho_c^*)^2\right)}-9  {f_u}^3 (-3+2 \rho_c^* ) {\phi}^{2/3}\right)\\ \nonumber 
\end{eqnarray}

\begin{equation}
 {A5}=2^{1/3}  {f_i}^3  {f_u}^3 \left(2592  {f_u}^2 \rho_c^* -67 {(2\phi)}^{1/3}\right)
\end{equation}

\begin{eqnarray}
 {B1}=278\ 2^{1/3}  {f_i}^4  {f_u}^4+3\ 2^{2/3} \sqrt{3} \sqrt{ {f_i}^4  {f_u}^6 \left(5  {f_i}^2+46  {f_i}  {f_u} \rho_c^*+243  {f_u}^2 (\rho_c^*)^2\right)}  {\phi}^{1/3}
\end{eqnarray}

\begin{eqnarray}
 {B2}= {f_i}^2  {f_u}^2 {\phi}^{1/3} \left(81\ 2^{2/3}  {f_u}^2 \rho_c^*-32 {\phi}^{1/3}\right)
 \end{eqnarray}

\begin{eqnarray}
 {B3}=12  {f_i} \left(8\ 2^{1/3} \sqrt{3}  {f_u} \sqrt{ {f_i}^4  {f_u}^6 \left(5  {f_i}^2+46  {f_i}  {f_u} \rho_c^*+243  {f_u}^2 (\rho_c^*)^2\right)}+9  {f_u}^3 \rho_c^* {\phi}^{2/3}\right)
\end{eqnarray}

\begin{equation}
 {B4}=2^{1/3}  {f_i}^3  {f_u}^3 \left(2592  {f_u}^2 \rho_c^*-67 {2\phi}^{1/3}\right)
\end{equation}

\begin{eqnarray}
 {B5}=12  {f_i}  \left(8\ 2^{1/3} \sqrt{3}  {f_u} \sqrt{ {f_i}^4  {f_u}^6 \left(5  {f_i}^2+46  {f_i}  {f_u} \rho_c^*+243  {f_u}^2 (\rho_c^*)^2\right)}-9  {f_u}^3 (-3+2 \rho_c^*) {\phi}^{2/3}\right) 
\end{eqnarray}
and
\begin{equation}
\phi=\left(23  {f_i}^3  {f_u}^3+243  {f_i}^2  {f_u}^4 \rho_c^*+9 \sqrt{3} \sqrt{ {f_i}^4  {f_u}^6 \left(5  {f_i}^2+46  {f_i}  {f_u} \rho_c^*+243  {f_u}^2 (\rho_c^*)^2\right)}\right)
\end{equation}

No analytical expression could be obtained for the surface separating LD-HD phases. For a given value of $\alpha$, $f_u$ and $f_i$, we obtained the corresponding $\rho_-$ and $J^{PBC}(\rho_-)$. On the LD-HD boundary, $J^{PBC}(\rho_-)$=$J^{PBC}(\rho_+)$. Hence, from the corresponding $\rho_+$, we calculated $\beta$ for given $f_u$ and $f_i$. In this way, points ($\alpha,\beta$) on the surface separating LD and HD phases were obtained for given $f_u$ and $f_i$.

\end{widetext}

\end{document}